\documentclass[final, a4paper]{aastex631}

\usepackage{color}
\usepackage{enumitem}
\usepackage{hyperref}
\usepackage{verbatim}
\usepackage{amsmath,amstext,mathrsfs}
\usepackage[all]{hypcap} 
\usepackage{afterpage}    
\usepackage{marginnote}
\usepackage{makecell}
\usepackage{subfigure}
\usepackage{multirow}

\shorttitle{Energy  Equipartition}
    \shortauthors{Zhao et al}
\usepackage{enumitem}
\setlist[enumerate]{listparindent=\parindent}
\makeatletter

\newcommand{\Rmnum}[1]{\expandafter\slowromancap\romannumeral #1@}
\makeatother

\begin{document}
    \hfuzz = 150pt
\title{\Large\bfseries Equation {\it vs.} AI: Predict Density and Measure Width of molecular clouds by Multiscale Decomposition}

\correspondingauthor{Guang-Xing Li, Keping Qiu}
\email{gxli@ynu.edu.cn,ligx.ngc7293@gmail.com,kpqiu@nju.edu.cn}

\author[0000-0003-0596-6608]{Mengke Zhao}
\affil{School of Astronomy and Space Science, Nanjing University, 163 Xianlin Avenue, Nanjing 210023, Jiangsu, People’s Republic of China}
\affil{Key Laboratory of Modern Astronomy and Astrophysics (Nanjing University), Ministry of Education, Nanjing 210023, Jiangsu, People’s Republic of China}

\author[0000-0003-3144-1952]{Guang-Xing Li}
\affil{South-Western Institute for Astronomy Research, Yunnan University, Kunming 650091, People’s Republic of China}

\author[0000-0001-6216-8931]{Duo Xu}
\affil{Department of Astronomy, University of Virginia, Charlottesville, VA 22904-4235, USA}
\affil{Canadian Institute for Theoretical Astrophysics, University of Toronto, 60 St. George Street, Toronto, ON M5S 3H8, Canada}

\author[0000-0002-5093-5088]{Keping Qiu}
\affil{School of Astronomy and Space Science, Nanjing University, 163 Xianlin Avenue, Nanjing 210023, Jiangsu, People’s Republic of China}
\affil{Key Laboratory of Modern Astronomy and Astrophysics (Nanjing University), Ministry of Education, Nanjing 210023, Jiangsu, People’s Republic of China}

\begin{abstract}

Interstellar medium is ubiquitous throughout the universe across multiple scales.
In this study, we introduce the {\it Multi-scale Decomposition Reconstruction} method, an equation-based model designed to derive width maps of interstellar medium structures and predict their volume density distribution on the plane of the sky from input column density data. 
This approach applies the {\it Constrained Diffusion Algorithm} \citep{2022ApJS..259...59L}, based on a simple yet common physical picture: as molecular clouds evolve to form stars, the density of interstellar medium increases while the scale decreases.
Validation on simulations confirms that this method accurately predicts volume density with minimal error. 
Notably, the equation-based model performs comparably or even more accurately than the AI-based DDPM model (Denoising Diffusion Probabilistic Models, \citealt{2023ApJ...950..146X}), which relies on vast parameters and substantial computational resources.
Unlike the "black-box" nature of AI, our equation-based model offers full transparency, making it easier to interpret, debug, and validate.
The simplicity, interpretability, and computational efficiency make it powerful not only for understanding complex astrophysical phenomena but also for complementing and enhancing AI-based methods.
    
\end{abstract}

\section{introduction}

The interstellar medium (ISM) plays a crucial role in a range of complex astronomical processes, including star formation, galaxy formation, and evolution \citep{2007ARA&A..45..565M,2016ARA&A..54..667S,2023ARA&A..61..561B}. 
It consists of gas and dust and is distributed on scales ranging from 10$^{4}$ to 10$^{-7}$ pc \citep{1987ARA&A..25...23S,2007ARA&A..45..565M,2015ARA&A..53..583H}. 
Understanding the multi-scale distribution of ISM is fundamental for addressing turbulence, energy spectra, energy cascades, and other physical processes in astronomy.

It is a well-established physical picture in astronomy that, as molecular clouds evolve to form stars, their density increases while their scale decreases \citep{1981MNRAS.194..809L,2018MNRAS.477.4951L,2019MNRAS.490.3061V}. 
In this context, the volume density is a fundamental parameter, directly governing the evolutionary timescale, the free-fall time ($t_{\rm ff} \approx ({\rm G}\rho)^{-\frac{1}{2}}$), and the critical thresholds for star formation. 
However, due to projection effects, the ISM is generally observed as surface density distributions on the plane of the sky (POS), where density and thickness (scale) information are degenerately mixed in the column density ($\Sigma \propto \rho \cdot r$). 
To overcome this inherent limitation, various methods have been developed, ranging from geometrical modeling, which assumes idealized shapes like cylinders or spheres \citep{2000MNRAS.311...85F,2016ApJ...822...10K}, to multi-tracer approaches that utilize chemical clocks or critical density excitation to probe the volume density, the 3D structural information \citep{2012ApJ...756...12L,2018A&A...610A..12B}. 
While effective in specific contexts, these methods often rely on strong morphological assumptions or require extensive, time-consuming multi-line observational campaigns to constrain the physical parameters.

A machine learning method, DDPM model(Denoising Diffusion Probabilistic Models, \citealt{2023ApJ...950..146X}), provides a powerful approach to predict volume density map from the column density of ISM.
This deep learning model employs a Denoising Diffusion Probabilistic Model (DDPM) trained on pairs of projected column density and projected volume density maps derived from Enzo magnetohydrodynamic (MHD) simulations of the ISM.
However, rely on deep learning introduces significant interpretability challenges.
Different to physical models, these AI models operate as "black box" with massive numbers of internal network parameters rather than explicit physical variables \citep{2023ApJ...950..146X}, making it difficult to explain its output or understand its decision-making processes. 
Astrophysical data often feature high dynamic range and significant uncertainty.
These characteristics pose major challenges for AI models, often leading to poor generalization and the generation of spurious artifacts.
In addition, training machine learning models costs massive computational resources.
Consequently, there is a distinct need for a mathematical model based on clear physical picture with minimal free parameters. 
Such a model facilitates the understanding of multi-scale processes and serves as a necessary complement to AI-based approaches.



To address this gap, we introduce a mathematical approach, {\it Multi-scale Decomposition Reconstruction}  (MDR), to predict the volume density distribution on the plane of sky using a multiscale decomposition algorithm, the {\it Constrained Diffusion Method} \citep{2022ApJS..259...59L}. 
Its core innovation lies in generating a \textit{spatial structural width map}, characteristic scale $l_{\rm c}$ from the column density $\Sigma$, which differs from the previous similar studies \citep{2016ApJ...822...10K,2020A&A...633A.132H,2025A&A...697A..20O}. 
Under the assumption of statistical isotropy, we apply the characteristic scale (width) to estimate the line-of-sight effective thickness ($l_{\rm t} \approx l_{\rm c}$), enabling volume density prediction through $⟨n({\rm H})⟩=\Sigma/l_{\rm t}$.
This approach extracts scale and density information directly from the $H_{2}$ column density. Unlike machine learning-based methods that rely on information from multiple tracers (combining continuum and spectral line emission) to determine ISM physical properties\citep{2018A&A...610A..12B}, our approach requires minimal input, only the column density map, while still producing reliable volume density estimates.
We compare the performance of this method with the DDPM to discuss strategies for enhancing AI applications in astronomy and the broader field of computer science.
The paper is organized as follows: Section\,\ref{sect2} details the MDR method. Section \ref{sec:validation} validates the method against MHD simulations (FLASH and Enzo) and discusses the isotropy assumption. Section\,\ref{sec4} compares the equation-based MDR model with the AI-based DDPM model. Finally, Section\,\ref{sec5} summarizes our findings.

\section{Multi-scale Decomposition Reconstruction}\label{sect2}

Star formation is a classical multi-scale physical process spanning over 10 orders of magnitude in density \citep{1987ARA&A..25...23S,2007ARA&A..45..565M}.
It is characterized by a hierarchical evolution driven by the interplay of multiple physical process, such as turbulence and self-gravity, where density increases as the scale decreases and gravitational forces eventually dominate the final collapse\citep{1981MNRAS.194..809L,2018MNRAS.477.4951L,2019MNRAS.490.3061V}.
Based on this astrophysical picture, we reconstruct the characteristic scale from the column density structure at each position on the plane of the sky (POS). 
By adopting an isotropic assumption, this characteristic scale is used to estimate the volume density, thereby enabling the study of star formation across multiple scales.
This method relies on the multi-scale decomposition to obtain the characteristic scale and predict the volume density in POS, called "Multi-scale Decomposition Reconstruction" (MDR).

The {\it Constrained Diffusion Method} \citep{2022ApJS..259...59L}, a multi-scale decomposition algorithm designed to decompose the total emission in astronomical data into multiple spatially dependent components. 
By iteratively identifying and isolating structural features, this method effectively separates local structures from the background across various physical scales. 
In this way, the dense structure within small scale, like core, filaments, and the diffuse one in large scale, like envelope, can be extracted from the complex column density map.
This method decomposes column density into multiple components with proper scales, where the intensity of each component corresponds to the column density contribution from structures at a specific spatial scale.
By computing a weighted average of the components at specific scales (MDR method), the characteristic scale obtained can be used to estimate the effective thickness along LOS and predict average volume density at POS.

\begin{figure}
    \centering
    \includegraphics[width=0.4\linewidth]{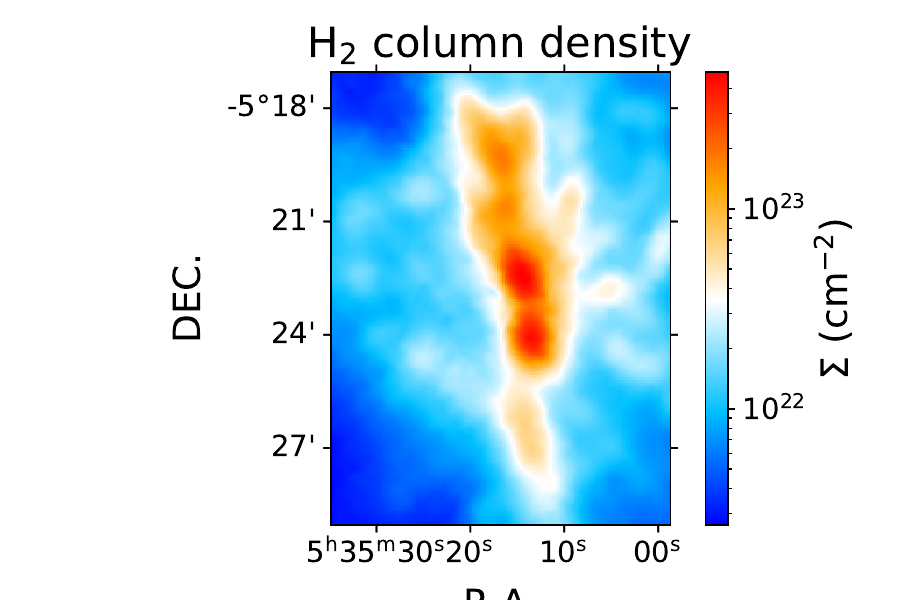}
    \includegraphics[width=0.95\linewidth]{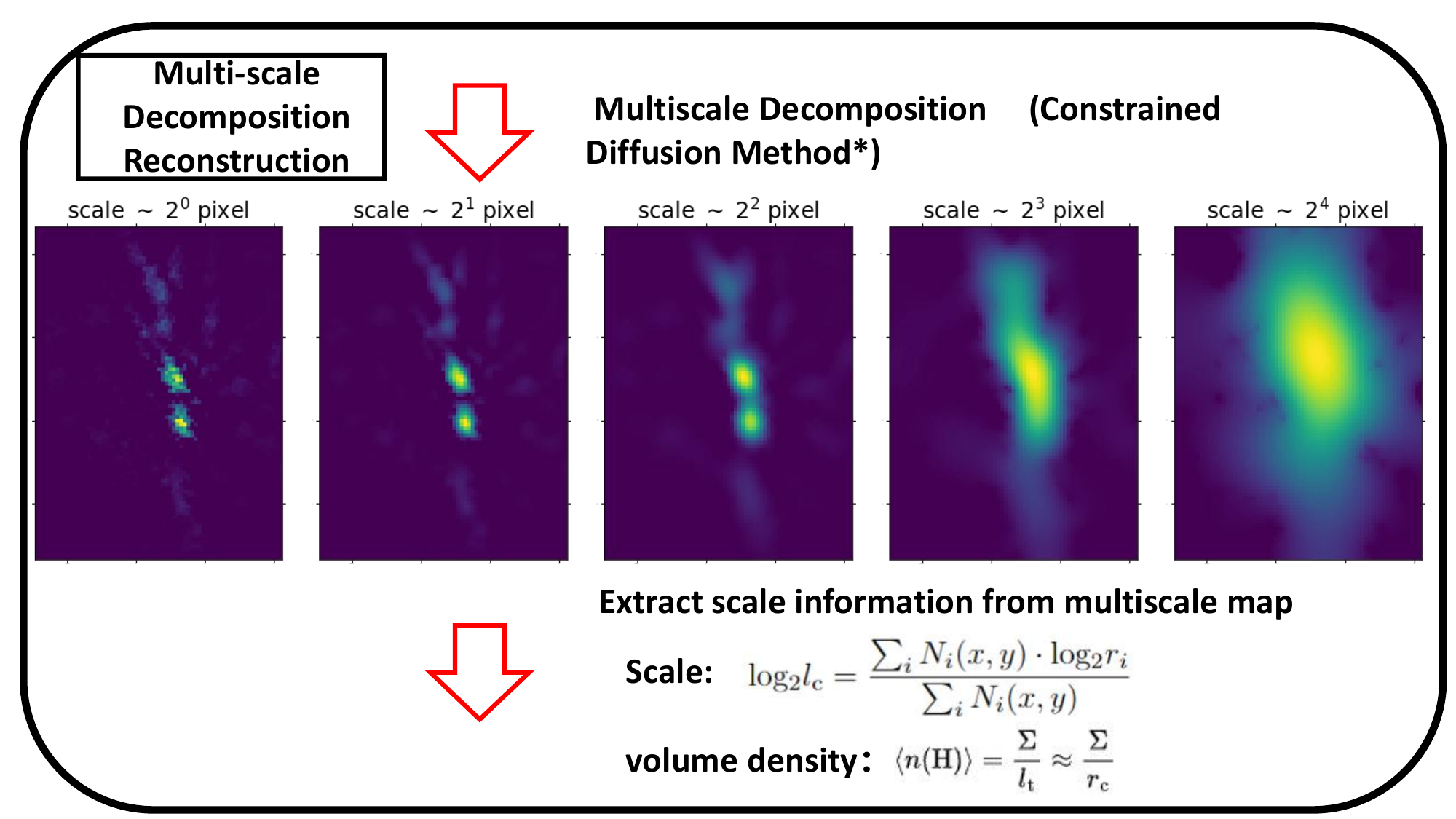}
    \includegraphics[width=0.8\linewidth]{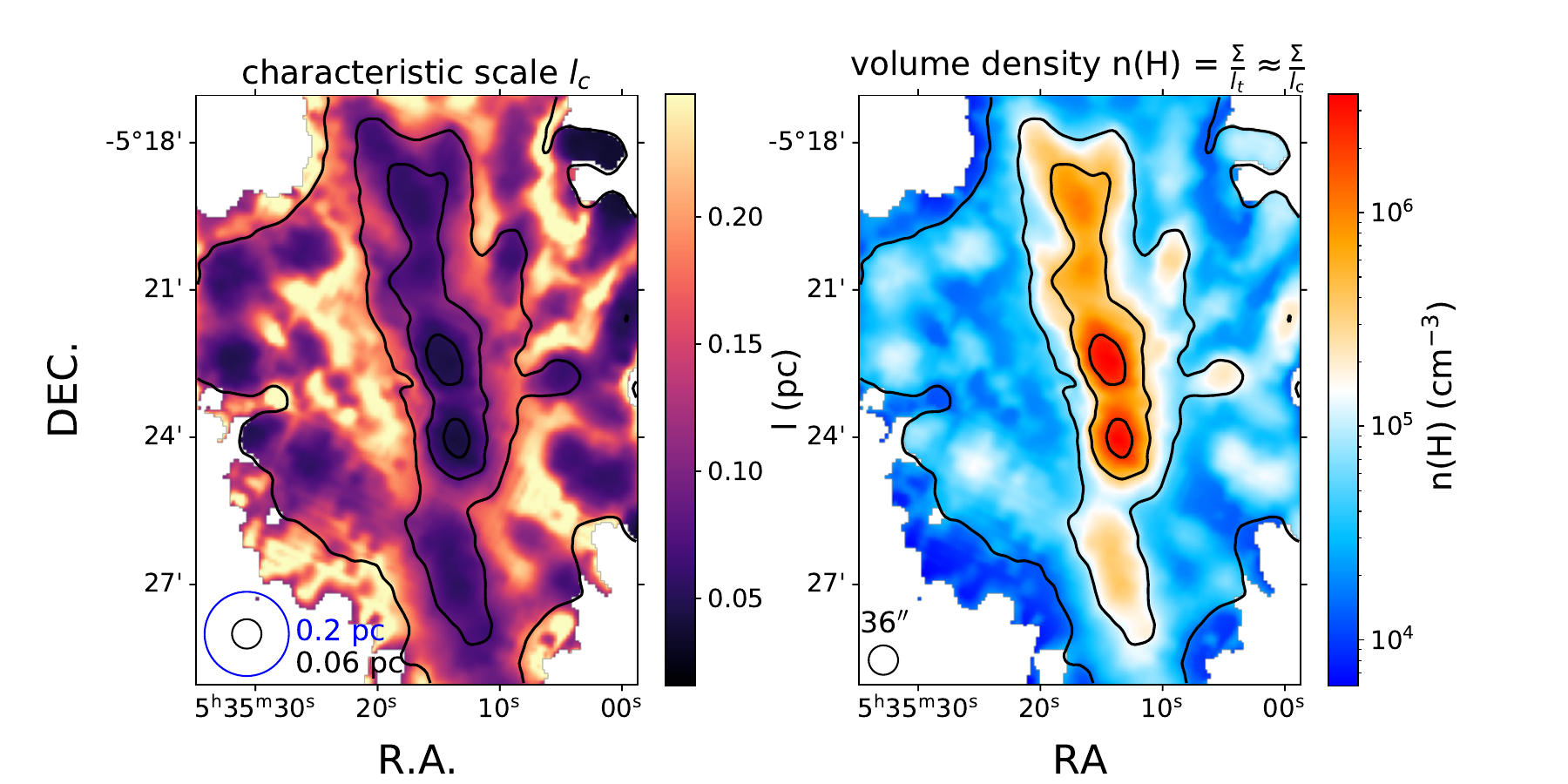}
    \caption{{\bf Pipeline of Multi-scale decomposition reconstruction to measure width map and predict volume density distribution.}
    The top panel shows the raw data of input, which is the column density of the molecular cloud, OMC-1.
    The middle panels present the calculated process of the Multi-scale Decomposition Reconstruction.
    The bottom panels are the output productions, width map and volume density distribution.}
    \label{figpipeline}
\end{figure}

\subsection{pipeline of Multi-scale Decomposition Reconstruction}

We introduce \textbf{Multi-scale Decomposition Reconstruction} (MDR): a novel methodology that reconstructs the characteristic scale of local emission, twice the distance from local position to structure center (width of structure).
Based on the isotropy assumption, where the effective cloud thickness is equivalent to the characteristic scale, the average volume density can be predicted in the plane of the sky. 
This approach consists of three sequential steps:

\begin{enumerate}
    \item \textbf{Characteristic scale measurement:} 
    The column density map is decomposed into structural components at logarithmically specific scales using the Constrained Diffusion Method \citep{2022ApJS..259...59L} (eg: specific scale size as  $2^n$ pixels, n=0, 1,..., n$_{\rm max}$, these specific scale sizes are the output of the Constrained Diffusion Algorithm). 
    Constrained Diffusion isolates the structures unique to each scale, resulting in a set of clean component maps. 
    Each component, $N_i$, contains structures of a specific size range, and summing these components reconstructs the original column density map.
    The characteristic scale $r(x,y)$ at each position is calculated as the intensity-weighted average in log-space:
    \begin{equation}\label{eqcr}
       {\rm log_2} l_{\rm c}= \frac{\sum_i N_i(x,y)\cdot {\rm log_2} r_i}{\sum_i N_i(x,y)} \,.
    \end{equation}
    This characteristic scale represents the effective spatial width of the local structures, such as the transverse width of a filament or the diameter of a clump or core, that dominate the column density emission at a given position.
    
    \item \textbf{Effective thickness estimation:}
    Assuming statistical isotropy of molecular cloud, the effective thickness along the LOS is estimated by characteristic scale at POS.
    \begin{equation}
        l_{\rm t} \approx l_{\rm c} \,.
    \end{equation}
    
    \item \textbf{Volume density prediction:}
    The average volume density $\langle n(\mathrm{H})\rangle$ is then derived by combining column density $\Sigma$ with the computed effective thickness:
    \begin{equation}\label{eqdensity}
        \langle n(\mathrm{H})\rangle = \frac{\Sigma}{l_{\rm t}} \approx \frac{\Sigma}{l_{\rm c}} \,.
    \end{equation}
    This represents the mass-weighted average density along the line of sight, physically interpreted as the uniform density required in a cloud of thickness $l_t$ to produce the observed column density $\Sigma$.
\end{enumerate}

Figure\,\ref{figpipeline} illustrates the complete MDR workflow applied to observational data. 
As demonstrated for the OMC-1 molecular cloud, the method first decomposes structures across scales, 
extracts width information through Eq.~\ref{eqcr}, and finally predicts volume density via Eq.~\ref{eqdensity}. 
The resulting characteristic scale maps and average volume density distributions directly correspond to local structure (dense clumps, filaments, diffuse gas, etc.) identified in the column density field.

\subsection{Measurement of Characteristic Scales}.

The core function of MDR is the measurement of characteristic scale (spatial structure width)  from a 2D astronomy image, such as H$_2$ column density, which first decompose the 2D image as several components then weights them.

The characteristic scale at each position in the plane of the sky is determined through a multi-scale decomposition of the column density map using the Constrained Diffusion Method.
This technique generates a series of intensity components (N$_i$(x,y)) at different specific size ($r_i$ as 2$^0$, 2$^1$, 2$^2$,..., pixel sizes, the output from Constrained Diffusion code). 
In general, scale structures of ISM distribute as power law \citep{1981MNRAS.194..809L,2018MNRAS.477.4951L,2022MNRAS.514L..16L}.
The characteristic scale is calculated as the intensity-weighted average of the logarithm base 2 of the scales (see Eq.\,\ref{eqcr})
Scales contributing more significantly to the observed column density structure at a given position naturally receive higher weight in the average calculation. 
Using log$_2$ r transforms the inherently power-law-like density distribution of the ISM (where density often scales inversely with size) into a linear space. 
This prevents larger spatial scales from disproportionately dominating the average, effectively mitigating contamination from large-scale structures. 
Physically, it represents the most characteristic spatial scale of the ISM structure that contributes to the emission at the POS position.

\subsection{Statistical Isotropy Assumption}

As indicated above, the core assumption of MDR is the statistical isotropy of molecular clouds on local scales, meaning that the structural width $l_c$ measured in the plane-of-sky (POS) approximates the effective thickness $l_t$ along the line-of-sight (LOS). 
The real molecular clouds exhibit complex structures and can be anisotropic, for instance, in the presence of strong magnetic fields that create sheet-like or filamentary structures \citep{2001ApJ...546..980O,2012A&ARv..20...55H,2025MNRAS.542.3246L}.
However, MDR is a necessary and effective simplification in the absence of direct 3D information. 
Its justification rests on the following grounds:
\begin{itemize}
    \item Physical Motivation: Turbulence theory suggests that the velocity and density fields are statistically isotropic in the inertial range \citep{2004ApJ...616..943L}
    \item Empirical Validation: We assess the implications of this assumption in Section\,\ref{sec:validation} through a series of tests.
    Our analysis of MHD simulations (see Section\,\ref{sec3.2}) reveals that the predicted volume density deviates from the ground truth by only a factor of 2. 
    In Section\,\ref{sec3.3}, we will directly compare the effective thickness $l_t$ derived from column density with the FWHM of the LOS density profile from the simulation, finding remarkable agreement. 
    These results strongly indicate that our method provides a robust estimate of the effective thickness in a statistical and averaged sense, despite localized anisotropies.
    \item Complementarity with AI: MDR method shows more consistent performance than the AI-based model in diffuse regions and extended dense structures (see Section\,\ref{sec4}). 
    This suggests that a clear, scale-based assumption can, in certain scenarios, be more robust than a complex AI model trained on simulations that may inherit their biases.
\end{itemize}

The MDR method presents a mathematically consistent framework for deriving spatially resolved structural width and average volume density maps estimated from column density maps.
The statistical isotropy of molecular clouds on local scales allows to MDR work well at most scales and extract 3D physical information from 2D observations as accurate approximation.
The forthcoming validation against MHD simulations (Sect.\,\ref{sec:validation}) will critically test its assumption and accuracy.
Appendix\,\ref{Ap.C} and \ref{Ap.D} present examples of how to apply MDR to actual observations.


\section{Validation in Observations and MHD simulations}\label{sec:validation}

To validate the proposed method, the reliability of MDR method is evaluated through four critical tests: 
(1) morphological consistency between derived characteristic scales and local structural features in observed column density maps; 
(2) quantitative comparison of predicted versus true volume density in MHD simulations; 
(3) geometric verification of effective thickness against simulated LOS density profiles ($z$-axis);
(4) statistical agreement between reconstructed and true density probability distribution functions (PDFs).
In addition, a comparison of MDR with molecular tracers in actual observation serves as another consistency check, as shown in Ap.\,\ref{Ap.E}.

\begin{figure}
    \centering
    \includegraphics[width=0.95\linewidth]{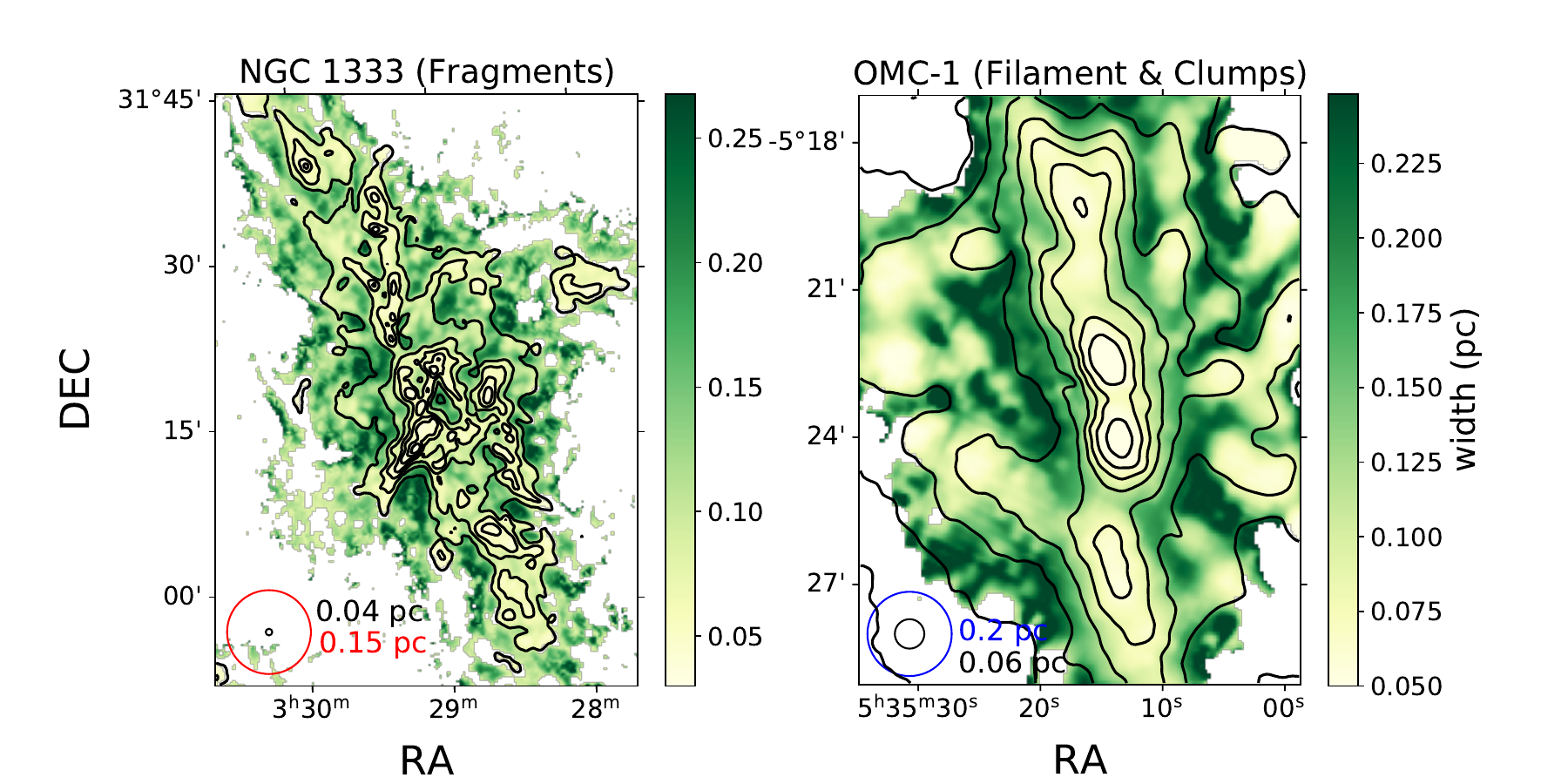}
    \caption{{Width map of molecular clouds NGC 1333 and OMC-1}.
    The panels present the width distribution at the plane of sky in molecular clouds NCG 1333 (left) and OMC-1 (right), which is equal to twice the distance from the local position to the structure center.
    The black contours show the distribution of raw data, column density from Herschel Gould Belt Survey, whose levels are from 10$^{21.5}$ to 10$^{23.5}$ cm$^{-2}$ with steps as 10$^{0.25}$.
    Scale annotations (circles) show representative large (a visual guide, e.g. 0.15 pc) and small (spatial resolution) scales for comparing the derived widths with physical ISM structures. }
    \label{figwidthmap}
\end{figure}

\subsection{Characteristic scale {\it vs.} structural width in observations}

In this section, we present a morphological comparison between the derived characteristic scales and the local structural features identified in the observed column density maps. 
Defining precise boundaries for physical structures is inherently difficult due to the hierarchical, multi-scale nature of the ISM \citep{2021A&A...653A.157L}, which limits the feasibility of direct quantitative matching.
We validate the characteristic scale measurements against observed structural widths as a morphological consistency check in two representative star-forming regions: a part of the integral-shaped filament in OMC-1 (dominated by dense clumps) and the fragmented cloud NGC 1333 (details shown in Ap.\,\ref{data}). 
Spatial resolutions were determined from observational beam sizes ($\sim$36$''$) and distances ($d_{\rm Perseus} \approx 270$ pc \citealt{2020A&A...633A..51Z}, $d_{\rm Orion\,A} \approx 400$ pc \citealt{2007A&A...474..515M}).

Figure\,\ref{figwidthmap} demonstrates a general morphological agreement between the derived characteristic scales and the physical structure widths across diverse morphologies. 
In OMC-1, the distribution of characteristic scales matches the width of the filament and the sizes of dense clumps. 
The red circle highlights a representative diffuse region where the extracted characteristic scale corresponds to the large-scale envelope.
Similarly, in the fragmented region of NGC 1333, the characteristic scales correspond well with the widths of massive dense cores.

Overall, the characteristic scales in both regions tend to decrease with increasing column density, indicating a negative correlation between scale and column density in the POS distribution. 
Generally, dense clumps and cores exhibit narrow scales while diffuse gas shows larger scales, aligned with the hierarchical physical picture \citep{1981MNRAS.194..809L,2018MNRAS.477.4951L,2022MNRAS.514L..16L}. 
We note that small characteristic scales are occasionally identified in lower density regions (e.g., at the edges of OMC-1). 
While some may trace local fragmentation, others likely represent algorithm artifacts arising from edge effects or the lack of structural constraints in diffuse, extended environments.
Nevertheless, this global correlation, valid across both extended (OMC-1) and fragmented (NGC 1333) environments, supports the robustness of the MDR method in extracting physical widths from 2D column density maps.



\begin{figure}
    \centering
    \includegraphics[width=0.95\linewidth]{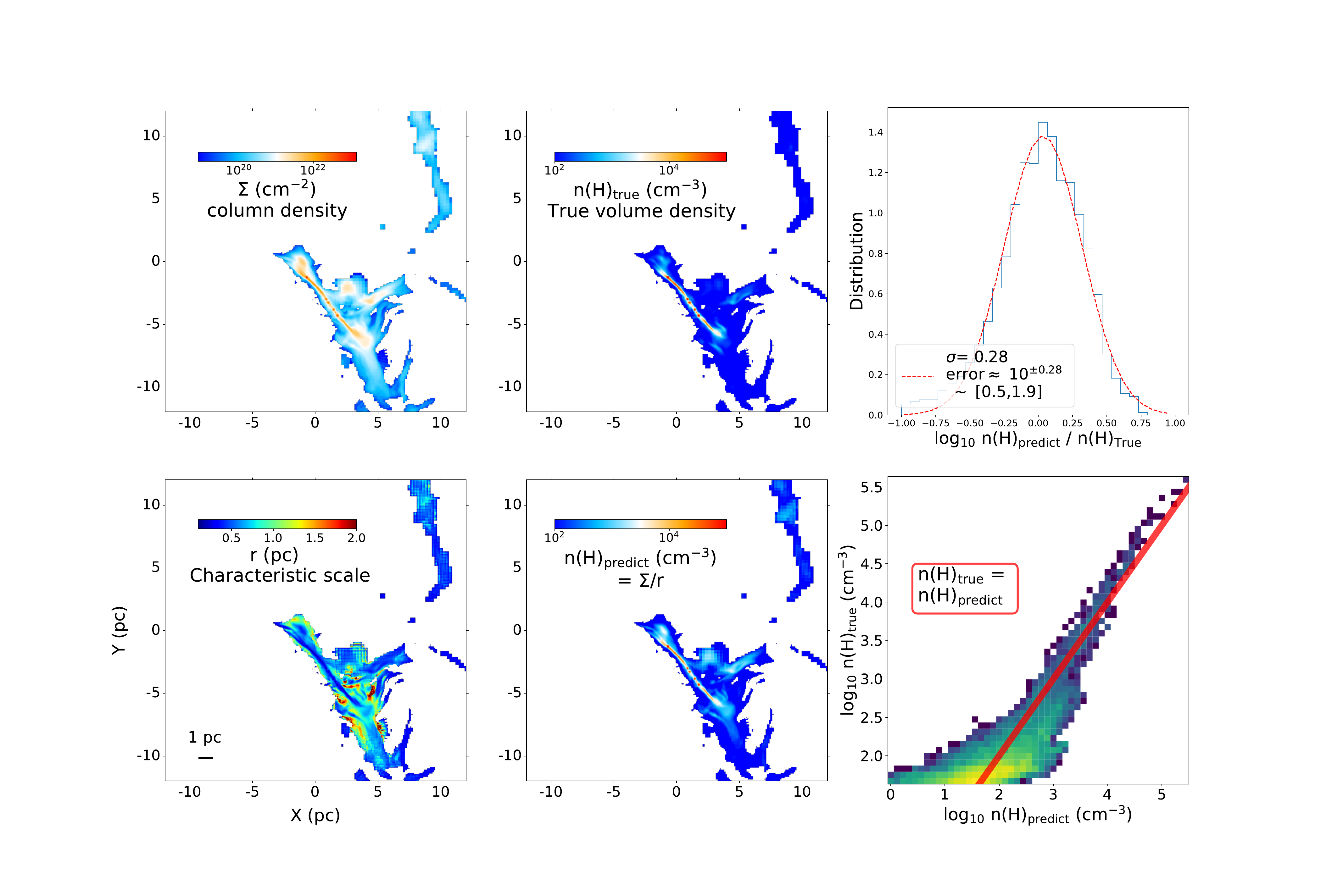}
    \caption{{\bf Distribution between the characteristic scale, predicted volume density, and true volume density in {\it FLASH} simulation filament clouds, M3}.
    The top panels show the H$_2$ column density and true volume density computed by 3D density structure of MHD simulation.
    The middle panels display the characteristic scale r derived by H$_2$ column density (see Eq.\ref{eqcr}), predict volume density (see Eq.\,\ref{eqdensity}), and the distribution between pseudo volume density and true volume density.
    The red line shows the n(H)$_{\rm predict}$ = n(H)$_{\rm true}$.
    The bottom panels present the predicted volume density map and the distribution between predicted volume density and true volume density with a binning size of 50.
    The red line means n(H)$_{\rm predict}$ = n(H)$_{\rm true}$.
    The histograms are constructed with a constant bin width of $\Delta \log_{10} (n) = 0.067$ dex
    This figure compares predicted volume density (MDR) and projected volume density along the z-axis, which the result of other projected directions is shown in Fig.\,\ref{figM3xy}.
    }
    \label{figM3}
\end{figure}

\begin{figure}
    \centering
    \includegraphics[width=0.95\linewidth]{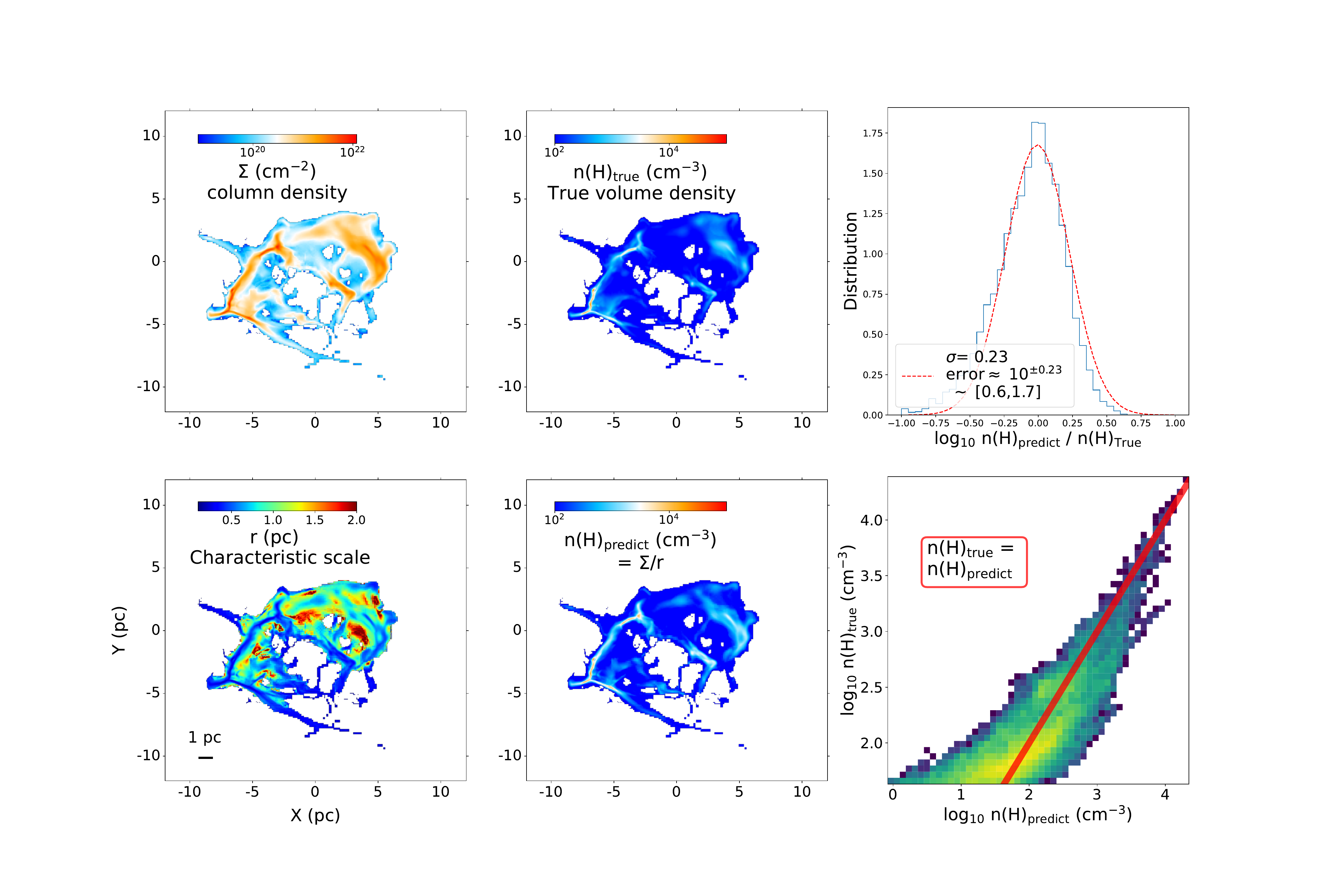}
    \caption{{\bf Distribution between the characteristic scale, predicted volume density, and true volume density in {\it Flash} simulation, extended clouds, M4}.
    They are the same as Fig.\,\ref{figM3} but for cloud, M4 with an extended structure.
    The comparison result along with other projections is shown in Fig.\,\ref{figM4xy}.}
    \label{figM4}
\end{figure}

\begin{figure}
    \centering
    \includegraphics[width=\linewidth]{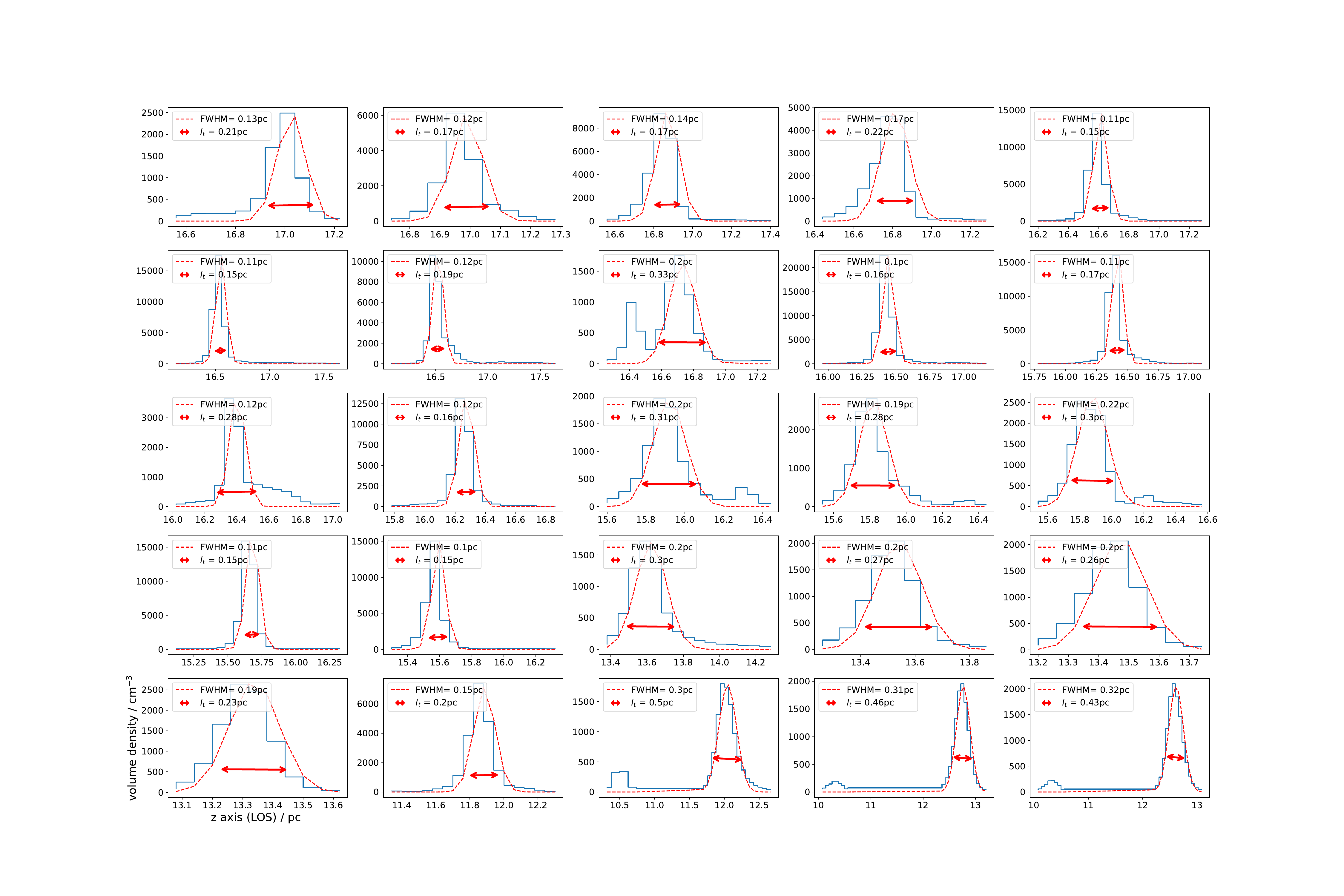}
    \caption{{\bf Effective thickness vs. density profile along the LOS in MHD simulation.}
    The blue lines show the density profiles along the LOS ($z$-axis) direction for the M4 cloud, which exhibits complex structure containing mixed clumps and filaments. 
    Red arrows present the effective thickness $l_{\rm t}$ derived from column density using Multi-scale Decomposition Reconstruction. 
    Dashed red curves show Gaussian fits to the density profiles, from which FWHM sizes are extracted.}
    \label{figZprofile}
\end{figure}

\subsection{Accurate measurement of scale and prediction of volume density}\label{sec3.2}

Extracting scale information directly from density structures, whether in 2D observational planes or 3D MHD simulations, is challenging because it is difficult to define precise boundaries for well-defined physical structures due to the multi-scale, complex shape of the ISM \citep{2021A&A...653A.157L}.
To bridge the gap between 2D observations and 3D physics, we rely on the assumption of statistical isotropy, positing that the characteristic scale measured in the plane of the sky ($l_c$) approximates the effective thickness along the line of sight ($l_t$).

To evaluate the accuracy of characteristic scale measurements and the resulting volume density predictions in ISM, we compare the predicted volume density with the true volume density in two MHD simulations conducted using the $FLASH$ code \citep{2017ApJ...850...62I,2019A&A...630A..97C,2020ApJ...905...14B}. 
The M3 cloud mainly exhibits filamentary structures, while the M4 cloud exhibits a complex, extended morphology with numerous substructures, such as filaments and clumps (detail shown in Ap.\,\ref{data}). 

To establish a valid ground truth for comparison, we must consider the physical nature of column density. 
Since column density represents the integration of mass along the line of sight, the corresponding "average" volume density is physically equivalent to the mass-weighted average density \citep{2001ApJ...546..980O,2023ApJ...950..146X}. 
We calculate this ground truth for the simulation data as:
\begin{equation}
\label{eq:ground_truth}
\langle n({\rm H}) \rangle_{\rm true} = \frac{\sum_{\rm LOS} n({\rm H})_i^2 \Delta V}{\sum_{\rm LOS} n({\rm H})_i \Delta V} ,,
\end{equation}
where $n({\rm H})_i$ is the number density and $\Delta V$ is the volume of each voxel along the line of sight. 
The mass-weighted average density correctly weights regions by their molecular content, especially in high-density regions \citep{2001ApJ...546..980O,2023ApJ...950..146X}.

This approach, MDR method, successfully predicts volume densities that closely match the true distribution for both M3 (single filamentary structure) and M4 (multi-clouds structure) in the FLASH simulations (see Fig.\ref{figM3} and Fig.\ref{figM4}). 
To verify robustness against viewing angles, we analyzed projections along the x, y, and z axes (detail of x, y axes shown in Fig.\,\ref{figM3xy} and \ref{figM4xy}).
The ratio between predicted and true volume densities follows a Gaussian distribution centered at unity (10$^0$). 
This statistical behavior indicates that the estimation errors are random rather than systematic, allowing us to quantify the uncertainty using the standard deviation. 
The resulting dispersions are 0.23-0.28 dex for M3 and 0.21-0.23 dex for M4 across different projections. 
Since the volume density is derived directly from the thickness estimate ($n \approx \Sigma/l_c$), this uncertainty directly reflects the precision of the effective thickness determination.

The MDR method achieves an accuracy of approximately $\pm$0.25 dex (a factor of $\sim$2), meaning that the MDR predictions generally fall within 0.5–1.9 times the ground truth. 
The increased dispersion in relatively diffuse regions is primarily attributed to the complex, volume-filling nature of large-scale structures. 
Unlike compact dense cores, which are spatially distinct, diffuse gas often exhibits extended or hollow density structures that overlap significantly along the line of sight (LOS). 
This superposition makes it inherently difficult to distinguish the true 3D spatial extent from 2D observations. 
Therefore, the global accuracy ($\sim$ 0.25 dex) represents a conservative estimate; in high-density regions critical for star formation, the method achieves significantly higher precision due to reduced projection confusion and tighter geometric constraints.


\subsection{Validation of Effective Thickness against Simulated LOS Density Profiles}\label{sec3.3}

To verify the effective thickness measured from MDR method, we analyzed 25 random positions from the high-density region of M4 (above $10^3$ cm$^{-3}$) simulation with complex ISM structure (see fig.\ref{figZprofile}). 
In these positions, we extracted the true density profile along the line-of-sight (z-axis) and plotted the characteristic scale calculated by column density (see Fig.\,\ref{figM4}as the effective thickness.
As Fig.\,\ref{figZprofile} shows, the characteristic scale as an effective thickness is aligned with density profile along the LOS direction.
To quantitatively analyze, we fitted a Gaussian function to the density profile to compute its FWHM width.
As shown in Fig.\,\ref{figZprofile}, the estimated characteristic scale ($l_c$) effectively captures the spatial extent of the cloud, showing strong consistency with the intrinsic density profile along the LOS (which is well-reproduced by a Gaussian fit).
On average, the MDR method based on the isotropy assumption could remain valid for deriving thickness estimates of realistic ISM.

Uncertainties may arise along lines of sight containing multiple distinct velocity components or spatial structures (superposition). 
Due to the high density contrast characteristic of the ISM, the projected column density is often dominated by the single most massive structure along the line of sight. 
Since the MDR method utilizes intensity-weighted averaging, the derived characteristic scale is primarily determined by this dominant high-density component rather than being a simple linear average of all structures. 
As shown in Fig.\,\ref{figZprofile}, the estimated values effectively capture the spatial extent of the principal emitting structure, proving that the characteristic length provides a robust measurement of the cloud's effective thickness even in the presence of projection effects. 


\begin{figure}
    \centering
    \includegraphics[width=0.5\linewidth]{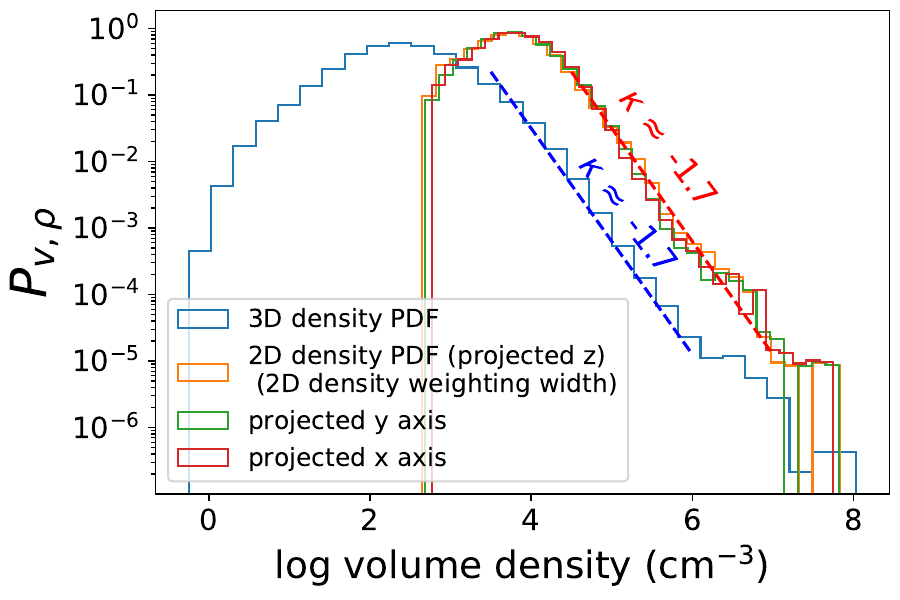}
    \caption{{A comparison of the density PDF in 3D space and reconstructed in 2D plane.}
    The light blue line shows the density PDF in 3D weight from 256$^3$ data points.
    The orange line displays the density PDF in the 2D plane weighting by thickness, equal to width, which both come from the column density distribution in 2D plane (256$^2$ pixels) using {\it Multi-scales Decomposition Reconstruction}.
    The blue and red present the fitting lines of power law distribution of density PDF from 3D space and rebuilt in 2D plane, respectively, which have the same slope ($\approx$ -1.7). }
    \label{figdensityPDF}
\end{figure}

\subsection{Density Probability Distribution Function}

The density probability distribution function (PDF) serves as a critical validation tool for MDR method in recovering intrinsic 3D statistics.
This fundamental diagnostic characterizes molecular cloud physics, distinguishing turbulent regimes (log-normal distribution) from gravitational collapse (power-law tail) \citep{2000ApJ...535..869K,2013ApJ...763...51F,2025A&A...696A..20G}. 
The density PDF must be volume-weighted, as it represents the fractional volume occupied by gas at each density level, information essential for understanding mass distribution.

A key observational limitation persists: while 2D volume density maps can be derived, true 3D density PDFs are difficult to obtain due to missing volume-fraction information across densities. 
MDR resolves this by reconstructing the volume dimension through thickness maps ($l_{t}$), derived under isotropy. 
By providing the effective thickness, MDR assigns each POS position an equivalent volume:  
$\Delta V = \Delta A \cdot l_t$  
where $\Delta A$ is the pixel area. This enables 3D density PDF reconstruction from observations.
We validate this approach using an Enzo simulation ($\beta_0 = 20$, $t = 0.6 t_{\rm ff}$) that realistically models complex molecular clouds \citep{2024ApJ...976..209Z} (details are shown in Ap.\,\ref{data}). 
As Fig.\,\ref{figdensityPDF} shows, we compare the true volume-weighted 3D PDF ($256^3$ voxels) with the MDR-derived PDF ($256^2$ pixels, the average volume density along LOS).
The reconstructed PDF matches key features of the true volume-weighted 3D distribution.
Their high-density regions exhibit consistent power-law slopes, demonstrating accurate recovery of collapse signatures.

It is important to note the apparent shift between the 3D and 2D PDFs in the low-density regime (see Fig.\,\ref{figdensityPDF}). 
This is not an algorithmic artifact but a natural consequence of projection. 
The 'True 3D PDF' samples every individual voxel, including the vast volumes of extremely low-density voids. 
In contrast, the MDR-derived 2D map represents the average volume density along the line of sight (LOS). 
This averaging process inevitably mixes high-density structures with low-density voids, thereby suppressing the low-density tail and compressing the dynamic range. 
Consequently, the peak of the log-normal distribution is shifted to higher values in the 2D representation. This phenomenon is inherent to the projection of turbulent media \citep{2001ApJ...546..980O,2025arXiv250917369L}. 

Despite this shift in the diffuse regime, the slopes of the high-density power-law tails, which trace gravitational collapse, remain consistent between the 3D ground truth and the MDR reconstruction, validating the method's ability to capture critical star formation physics.
In low-density regions, the MDR-derived PDF also exhibits the log-normal distribution with a reduced dynamic range.
The single value of average density and effective thickness in each position naturally compresses the full LOS density distribution, yet maintains the essential statistical properties of the turbulent ISM.
The threshold density in the 2D MDR-derived density PDF marks the transition between the power-law and log-normal regimes, corresponding to a critical average volume density along the line of sight.

The consistency between the true density PDF in 3D space and the reconstructed 2D version demonstrates the effectiveness of the isotropy-based reconstruction even for complex ISM structures.
The reconstruction fidelity stems from: (1) the MDR method's accurate decomposition of column density into volume density and thickness components, and (2) the fact that the projected column density maps retain statistically meaningful signatures of the 3D density structure.
In addition, MDR results are consistent with molecular tracer estimates within the latter's inherent uncertainties (detail shown in Ap.\,\ref{Ap.E}).
This confirms that the MDR method effectively extracts volumetric information from projected data.

\section{Equation-based Model vs AI}\label{sec4}

\begin{figure}
    \centering
    \includegraphics[width=0.95\linewidth]{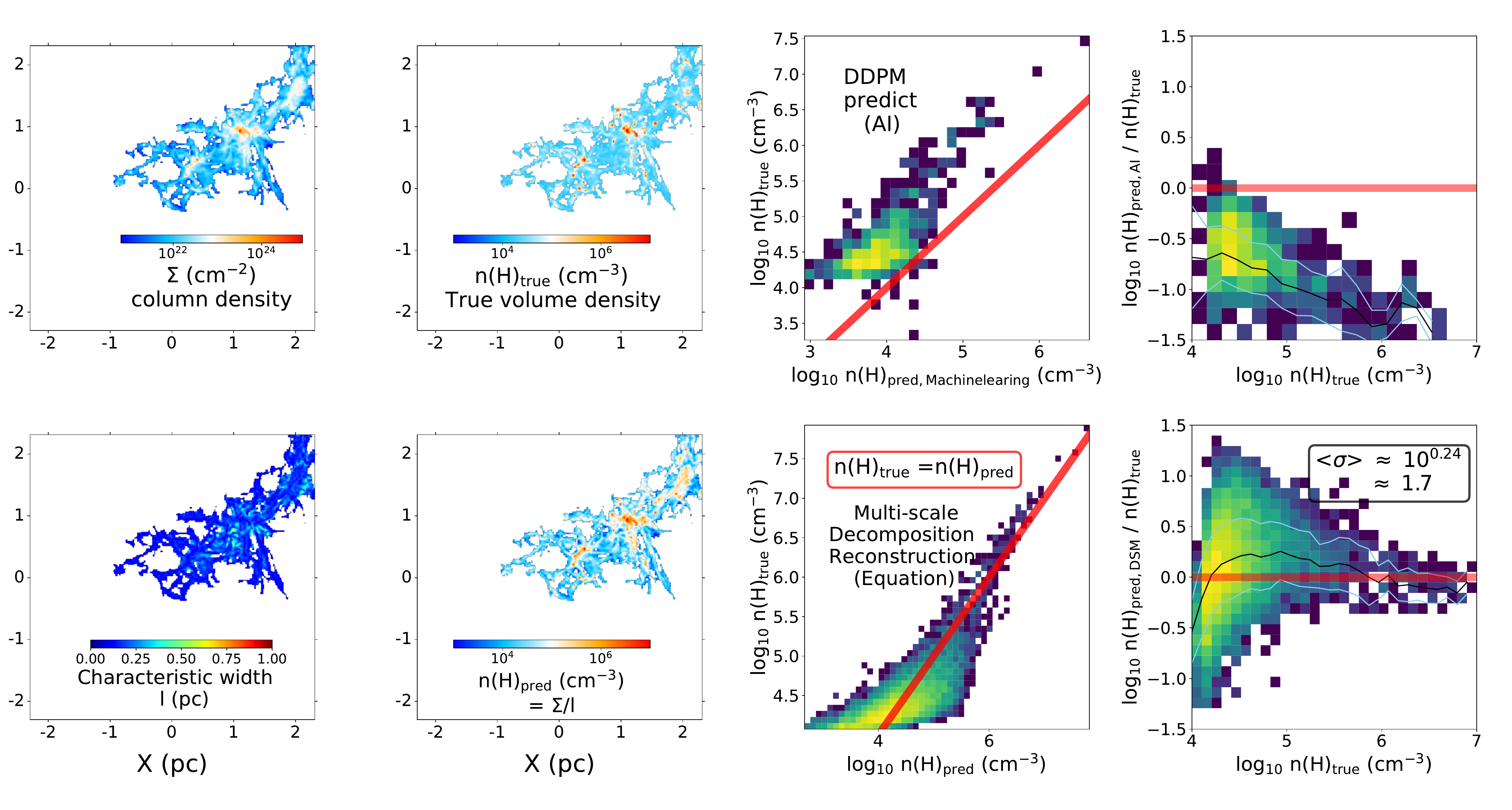}
    \caption{{\bf Distribution between the characteristic scale, predicted volume density, and true volume density in {\it Enzo} simulation of complex clouds}.
    They are the same as Fig.\,\ref{figM3} but for a complex cloud.
    The right-top panel shows the 2D histogram between true volume density from the 3D density structure and the predicted volume density computed by the machine learning algorithm, DDPM \citep{2023ApJ...950..146X}.
    The right panel shows the ratio between predicted volume density and true volume density.
    The red thin lines present the average ratio at each volume density range and the blue thin lines show the standard deviation of this distribution.}
    \label{figEnzo}
\end{figure}

\begin{figure}
    \centering
    \includegraphics[width=\linewidth]{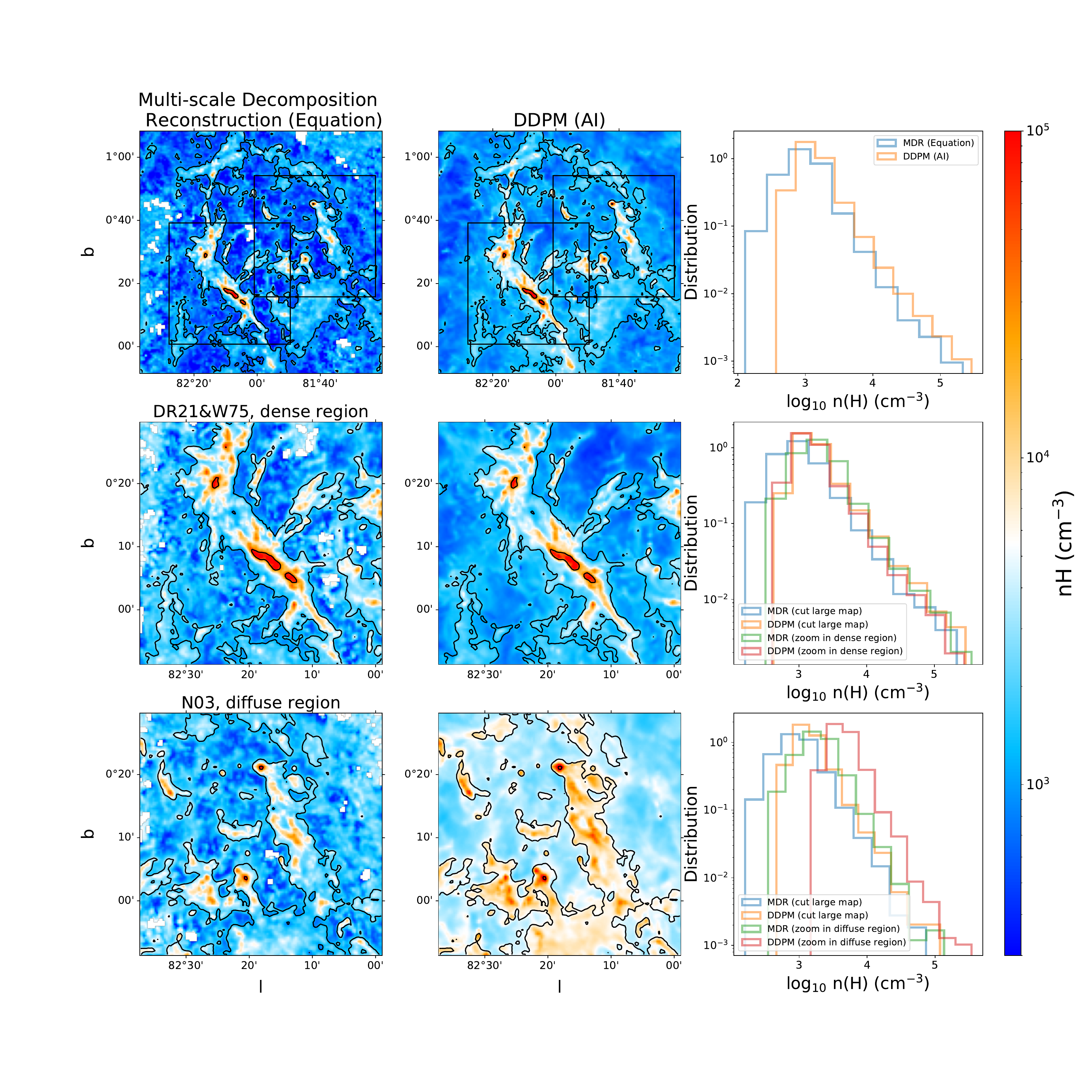}
    \caption{{\bf Predicted volume density in complex molecular cloud, Cygnus-X, using Multi-scale decomposition reconstruction (equation) and DDPM (AI).}
    The top left and middle panels display the predicted volume density map using equation-based model and AI model, respectively.
    The black contours show the structure of input data, the column density map of Cygnus-X, whose levels are 10$^{22}$ and 10$^{23}$ cm$^{-2}$. 
    The middle row shows a zoomed view of the dense region, and the bottom row displays the zoomed view of the diffuse region.
    The right panels in the middle/bottom rows show density distributions for the respective zoomed areas.
    }
    \label{figCyg}
\end{figure}

Machine learning (ML) algorithms are increasingly standard in astronomy, used to classify galaxies \citep{2017MNRAS.464.4463K}, detect exoplanets \citep{2018AJ....155...94S}, and predict ISM properties \citep{2023ApJ...942...95X,2023ApJ...950..146X}. 
However, as purely data-driven tools, these models face challenges in interpretability and generalization, particularly when applied to physical regimes outside their training distribution.

In this section, we compare the analytical MDR method with a representative generative AI model, the Denoising Diffusion Probabilistic Model (DDPM; detailed in \citealt{2023ApJ...950..146X}). 
Our goal is to contrast the robustness of a physics-based approach with a data-driven one. 
While AI-based models show promise, they are constrained by the quality, dynamic range, and structural diversity of their training data. 
We demonstrate how the analytical nature of MDR offers a reliable estimate of volume density and provides physical insight into the 'black-box' behavior of AI models, revealing specific structural biases in their predictions.

\subsection{Comparison with Data-Driven Approaches: Physics-based vs. Deep Learning}\label{sec4.1}

To evaluate model performance under realistic observational conditions, we analyze a molecular cloud from Enzo MHD simulations featuring intricate density structure (detail of Enzo simulation shown in Ap.\,\ref{ap.enzo}). 
Selected for its structural complexity yet minimal line-of-sight superposition, this target presents extended high-density structures and nested substructures that rigorously test both methods under conditions mirroring actual observations. 
The non-uniform density distribution of cloud specifically challenges the isotropy assumption of MDR while probing DDPM's susceptibility to training biases in scale recognition. 

Results demonstrate performance differences between the approaches (see Fig.\,\ref{figEnzo}). 
MDR shows robust agreement with ground truth, accurately capturing extended dense features through its scale-density framework with mean density ratios near unity and dispersion around 0.24 dex. 
In contrast, the DDPM exhibits systematic deviations, significantly underestimating the volume density across the board.
In low-density regions, the density predicted by DDPM is underestimated by around 0.5 dex, while in high-density regions, the underestimation reaches 1 dex. 

The performance gap could originate from fundamental methodological differences. 
MDR leverages first-principles physics, the statistical isotropy of the cloud and the correlation between scale and density, to extract structural information directly. 
Meanwhile, although the DDPM model covers a wide density range ($10^0 - 10^7 \text{ cm}^{-3}$, \citealt{2023ApJ...950..146X}), it shows larger deviations in recovering the correct densities for the complex structures in our test. 
This suggests that the discrepancy is likely not due to density limits, but rather a sensitivity to morphological priors.
The AI model appears to rely on a biased learned correlation between scale and density. 
In its training set, large-scale structures are predominantly diffuse, while small-scale features are typically dense cores. 
Consequently, when encountering the extended but high-density structures in our test, the model tends to interpret them as large, low-density clouds.
To investigate this hypothesis, we will continue to assess the robustness of these two models in the complexity target from actual observation.

\subsection{Studying Complex structure at POS from observation}

To further validate these models under complex observational conditions, we apply both the MDR method and DDPM to the Cygnus-X region \citep{2019ApJS..241....1C}. 
This region features a complex spatial hierarchy, including dense subregions such as DR21 and W75 \citep{2010A&A...520A..49S,2021ApJ...918L...4C,2024A&A...687A.207Z}, as well as fragmented, diffuse regions like DR17, N12, and N26 \citep{2016A&A...587A..74S,2019ApJS..241....1C,2024A&A...684A.142Z}.

The volume density distributions predicted by the equation-based MDR method and the AI-based DDPM are broadly consistent (see Fig.\,\ref{figCyg}). 
This agreement is largely driven by the high-density regions (e.g., DR21/W75), where both methods effectively capture the compact core structures. 
Since these dense cores dominate the statistical weight of the density distribution's high-end tail, the overall PDFs appear similar. 
Quantitatively, MDR produces values approximately 1.5 times ($\sim$ 0.18 dex) lower than DDPM, which remains well within the accepted uncertainty range.
These results demonstrate that the MDR method, relying on a simple physical framework, achieves performance comparable to the computationally intensive AI model in capturing the global statistics of star-forming regions.

However, a detailed multi-scale analysis reveals a critical divergence. 
When examining specific subregions at higher magnification ("zoom-in"), we observe distinct behaviors. 
In dense regions like DR21 and W75, both MDR and DDPM predict consistent volume densities regardless of whether they are viewed in the zoom-in map or within the larger field of view. 
This aligns with previous studies indicating that DDPM is robust for dense cores. 
In contrast, significant discrepancies emerge in the diffuse, fragmented region N03. 
Here, the DDPM's zoomed-in prediction yields significantly higher densities than both the MDR result and its own prediction derived from the large-scale map. 
Conversely, the MDR method demonstrates scale invariance, maintaining consistent density estimates before and after zooming. 
The observed deviations are small ($\lesssim$ 0.2 dex) and likely stem from the truncation of large-scale structural information at the boundaries.

\subsection{Complementary Strengths of MDR and DDPM}

The scale-dependent inconsistency in DDPM (Region N03) indicates that the AI model exhibits structural hallucination in diffuse environments. 
Because the AI-based model lacks large-scale morphological continuity, it appears to default to training priors, such as misinterpreting diffuse gas as dense cores. 
This error presents a limitation of data-driven approaches: they depend on synthetic priors that may not capture the full morphological complexity of real observations.

In comparison, the MDR method relies on a clear physical picture: the scale-density correlation and statistical isotropy. 
Since it is independent of training data, MDR maintains physical consistency across scales and avoids the generalization errors seen in the AI model.

MDR is also interpretable and efficient. 
Unlike the "black box" of deep learning, MDR can provide transparent intermediate steps (Fig.\,\ref{figpipeline}), which allow users to trace the origin of density estimates. 
It also runs efficiently on general-purpose CPUs, avoiding the high computational costs of generative AI (Appendix \ref{ApB}).

We suggest that equation-based models MDR complement AI rather than replace it. 
MDR serves as a physics-based framework that can provide benchmarks for AI interpretability and can guide the design of better training strategies, for example, by acting as a physical constraint in self-supervised learning. 
This combination of traditional mathematical derivations and modern machine learning provides a new way toward greater accuracy in astrophysics.

\section{Summary}\label{sec5}

Our results demonstrate that \textit{Multi-scale Decomposition Reconstruction} (MDR) reliably predicts volume density and structural width with $\sim$0.25 dex accuracy, outperforming or complementing AI-based methods in interpretability and computational efficiency.

The \textit{Constrained Diffusion Method} \citep{2022ApJS..259...59L} decomposes the column density map into multi-scale components. 
These column density components $N_i(x,y)$ at specific scales (eg: $r_i = 2^0, 2^1, 2^2, \ldots$ pixel size, the output of the constrained diffusion algorithm), where each component represents structural contributions at special scales.
Using the components with scales, the characteristic scale $r(x,y)$ at each pixel can be calculated through intensity-weighted averaging:
\begin{equation}
    \log_2 l_{\rm c} = \frac{\sum_i N_i(x,y) \cdot \log_2 r_i}{\sum_i N_i(x,y)} \,.
\end{equation}
where most relevant characteristic scale $l_{\rm c}$ can be extracted from small dense cores to large diffuse structures.
Under the statistical isotropy assumption (where the POS structural width approximates the LOS thickness), we equate the effective thickness to the characteristic scale ($l_t \approx l_c$).
Thus, the volume density is computed by column density $\Sigma$ and effective thickness $l_t$:
\begin{equation}
n(\mathrm{H})_{\rm pred} = \frac{\Sigma}{l_t} \approx  \frac{\Sigma}{l_{\rm c}} \,.
\end{equation}
In this work, we conducted four validation tests to evaluate the accuracy of the reconstruction and the robustness of the characteristic scale estimation: comparing the characteristic scale with observed density structure,  predicted versus true density in MHD simulation, the scale with density profile along LOS in simulation, and reconstructed density-PDF with true density-PDF in simulation.
Validated against Flash simulations, MDR achieves $\sim$0.25 dex accuracy (factor of 2 uncertainty) across diverse molecular cloud structures. 

{\it Multi-scale Decomposition Reconstruction} method, as a classical equation-based model, relies solely on a intuitive physical picture and basic assumptions to interpret the complex density structures across multiple scales in both the LOS (Enzo simulations) and the POS (Cygnus-X observations). 
MDR demonstrates consistent accuracy across diverse structural regimes. While AI (DDPM, \citealt{2023ApJ...950..146X}) performs comparably on compact dense cores, it shows significant instability on complex or extended structures (as seen in both simulations and diffuse observational regions). MDR's physics-based approach avoids these scale-dependent biases.
By applying a mathematical model based on fundamental physical principles and straightforward assumptions, the equation-based model significantly reduces the complexity involved in understanding multi-scale structures. 
In contrast, AI-based models, which depend on vast numbers of parameters to analyze these intricate structures, often involve opaque intermediate processes that lack clear physical interpretability.
Additionally, the {\it Constrained Diffusion Method} is lightweight, CPU-efficient (Ap.\,\ref{ApB}), and easily implemented in Python 3 (code address: \url{https://github.com/gxli/volume-density-mapper}).


In summary, the {\it Constrained Diffusion Method} demonstrates high accuracy and computational efficiency but also serves as a valuable tool for understanding biases in AI training sets and improving AI models.
Its transparent design and resource efficiency make it a promising complement to AI, especially in the study of astrophysical systems.

\bibliographystyle{aasjournal}
\bibliography{reference}

\appendix

\section{Data}\label{data}

\subsection{Flash Simulation}\label{ap.flash}

M3 and M4 clouds are three-dimensional numerical MHD simulations \citep{2017ApJ...850...62I,2019A&A...630A..97C,2020ApJ...905...14B} with self-gravitating, magnetized, SN-driven, turbulent, multiphase-ISM using the FLASH v4.2.2 adaptive mesh refinement code (AMR; \citealt{2000ApJS..131..273F}). 
The evolving time is around 3 Myrs, and the spatial resolution is 0.06 pc.
The M3 cloud has a classical filament structure and the M4 cloud with extended structure have multiple sub-structures like filament and clumps.

\subsection{Enzo Simulation}\label{ap.enzo}
For this study, we identified a molecular cloud through self-gravitating simulations using the constrained transport MHD option in Enzo (MHDCT) code \citep{2012ApJ...750...13C,2015ApJ...808...48B,2020ApJ...905...14B}. 
The simulations aimed to analyze the effects of self-gravity and magnetic fields on supersonic turbulence in iso-thermal molecular clouds. 
We use the simulation data ($\beta_0$ = 20, t $\approx$ 0.76 Myr), which initial conditions are:
\begin{equation}
    {\cal M}_s = \frac{v_{\rm rms}}{c_s} = 9\,.
\end{equation}
\begin{equation}
    \alpha_{vir} = \frac{5 v_{\rm rms}^2 }{ 3 G \rho_0 L_0^2} = 1\,.
\end{equation}
\begin{equation}\label{eq3}
    \beta_0 = \frac{8\pi c_s^2 \rho_0}{B_0^2} = 20\,.
\end{equation}
This simulation setup is chosen such that the results resemble the structure of an actual star-forming molecular cloud, which is super-Alfvenic and close to the high-density molecular cloud \citep{2024ApJ...976..209Z}.
The voxel resolution in the Enzo simulation is around 0.018\,pc, and its size is 256$^3$ voxels. 

\subsection{Column density map}

The H$_2$ column density map in the molecular cloud, Orion A, NGC\,1333 (see Fig.\,\ref{figOrionA}), comes from Herschel Gould Belt Survey \footnote{\url{http://www.herschel.fr/cea/gouldbelt/en/Phocea/Vie_des_labos/Ast/ast_visu.php?id_ast=66}}\citep{2010A&A...518L...2P,2010A&A...518L...3G,2013ApJ...763...55R,2013ApJ...777L..33P}, which apply the SED fitting procedure \citep{2015MNRAS.450.4043W} on the Herschel continuum at wavelengths of 70, 160, 250, 350, 500$\mu$m.
The resolution is around 36$''$.
The spatial resolution of Orion A and Perseus is 0.06 pc and 0.04 pc, respectively.

The H$_2$ column density map in the complex star-forming region Cygnus-X comes from \citealt{2019ApJS..241....1C}, which apply the SED fitting procedure \citep{2015MNRAS.450.4043W} on the Herschel continuum at wavelengths of 70, 160, 250, 350, 500$\mu$m and JCMT SCUBA-2 continuum data at 450 and 850 $\mu$m, IRAM 1.2 mm continuum images.
The resolution is around 18$''$ and the spatial resolution is around 0.12 pc.

\section{Computing Time estimate $\&$ Application of Observation}\label{ApB}

We have tested the MDR algorithm on several observational datasets to estimate its computational time in realistic application scenarios. 
The test data came from the Herschel Gould Belt Survey \footnote{\url{http://www.herschel.fr/cea/gouldbelt/en/Phocea/Vie_des_labos/Ast/ast_visu.php?id_ast=66}} and the Cygnus-X Survey\citep{2019ApJS..241....1C}.

The tests were conducted using a single core of a general-purpose CPU, specifically an AMD Ryzen 7 7700X. 
Although this CPU is not a top-tier model, such as the Intel i9 or AMD Ryzen 9, it provides a reliable baseline for performance evaluation. 
The estimated computing times are summarized in Tab.,\ref{tabCtime}.

For data sizes equal to or less than 1024$^2$ pixels, the computing time is remarkably short. 
However, as the data size increases, the computation time increases significantly. 
This increase is due to the diffusion equation \citep{2022ApJS..259...59L}, where the required time $t$ is proportional to the square of the characteristic scale $r^2$.

\begin{figure}
    \centering
    \includegraphics[width=0.95\linewidth]{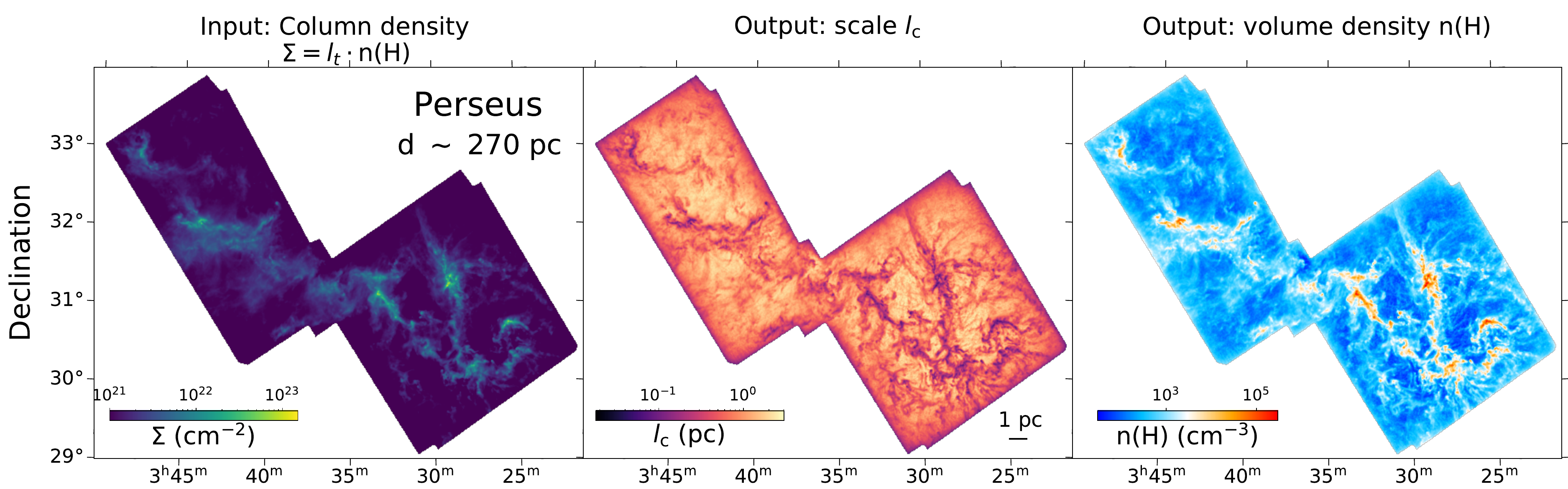}
    \includegraphics[width=0.95\linewidth]{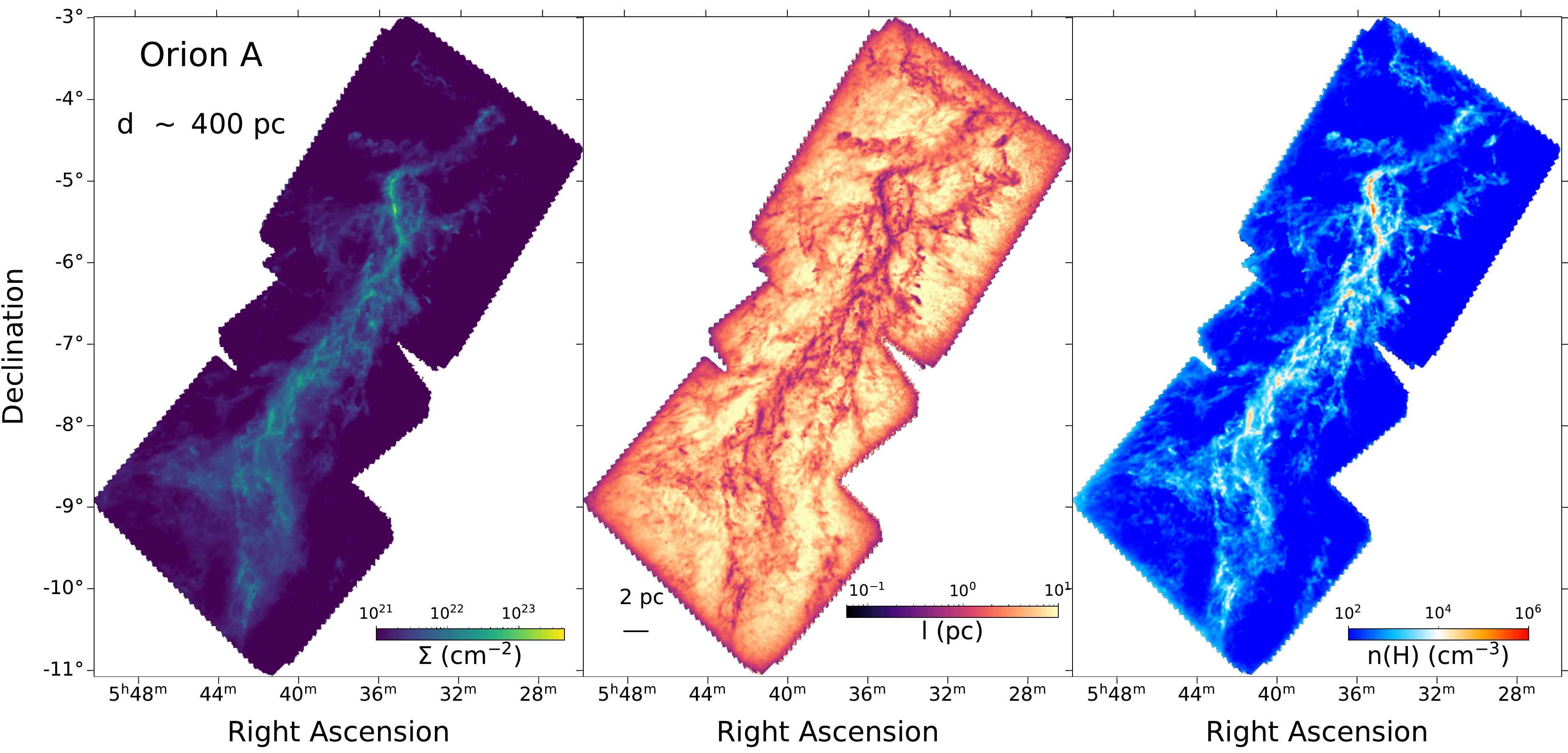}
    \includegraphics[width=0.95\linewidth]{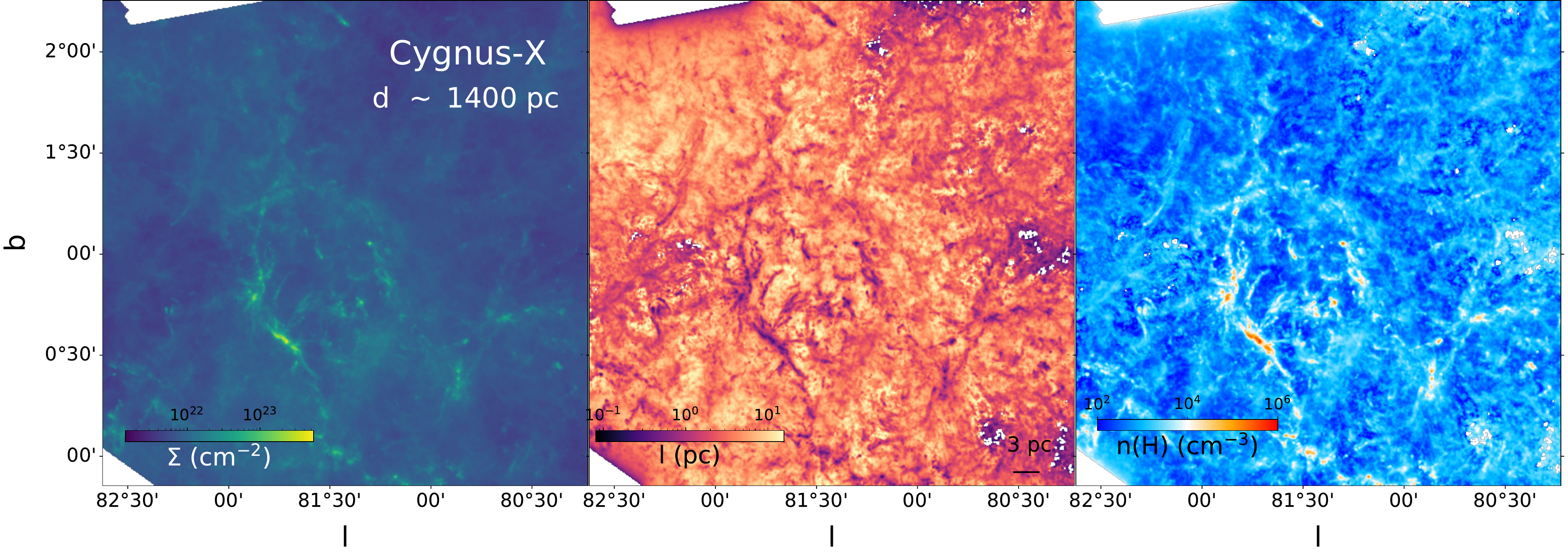}
    \caption{{Application of {\it multi-scale Decomposition Reconstruction} on Observations with Large cover region.}
    The top, middle, and bottom panels show the three observations, Perseus (distance $\sim$ 270\,pc), Orion\,A (distance $\sim$ 400\,pc) and Cygnus-X (distance $\sim$ 1400\,pc).
    The left panels display the column density map, the input of {\it multi-scale Decomposition Reconstruction}.
    The outputs are shown in the middle and right panels, the width map and volume density distribution of three molecular clouds.
    }
    \label{figOrionA}
\end{figure}

Fig.\,\ref{figOrionA} shows an example of the application of the MDR algorithm to observational data from the Orion A molecular cloud. Using the column density as input, the algorithm produces both the width map and the volume density distribution as outputs.

\begin{table}[h]
    \centering
    \caption{ Computing time estimated of multi-scale Decomposition Reconstruction algorithm in a single core of CPU (AMD Ryzen 7 7700X)}
    \begin{tabular}{c c}
    \hline
       Data size  & Computing time \\
    \hline
       128$^2$ pixels  &  $\sim$ 0.3 seconds \\
       256$^2$ pixels & $\sim$ 1 second \\        
       512$^2$ pixels & $\sim$ 14 seconds \\
       1024$^2$ pixels & $\sim$ 1 minute \\
    \hline
    \end{tabular}
    \label{tabCtime}
\end{table}

\section{Projected in other axis of M3 and M4} \label{Ap.C}

\begin{figure}
    \centering
    \includegraphics[width=0.95\linewidth]{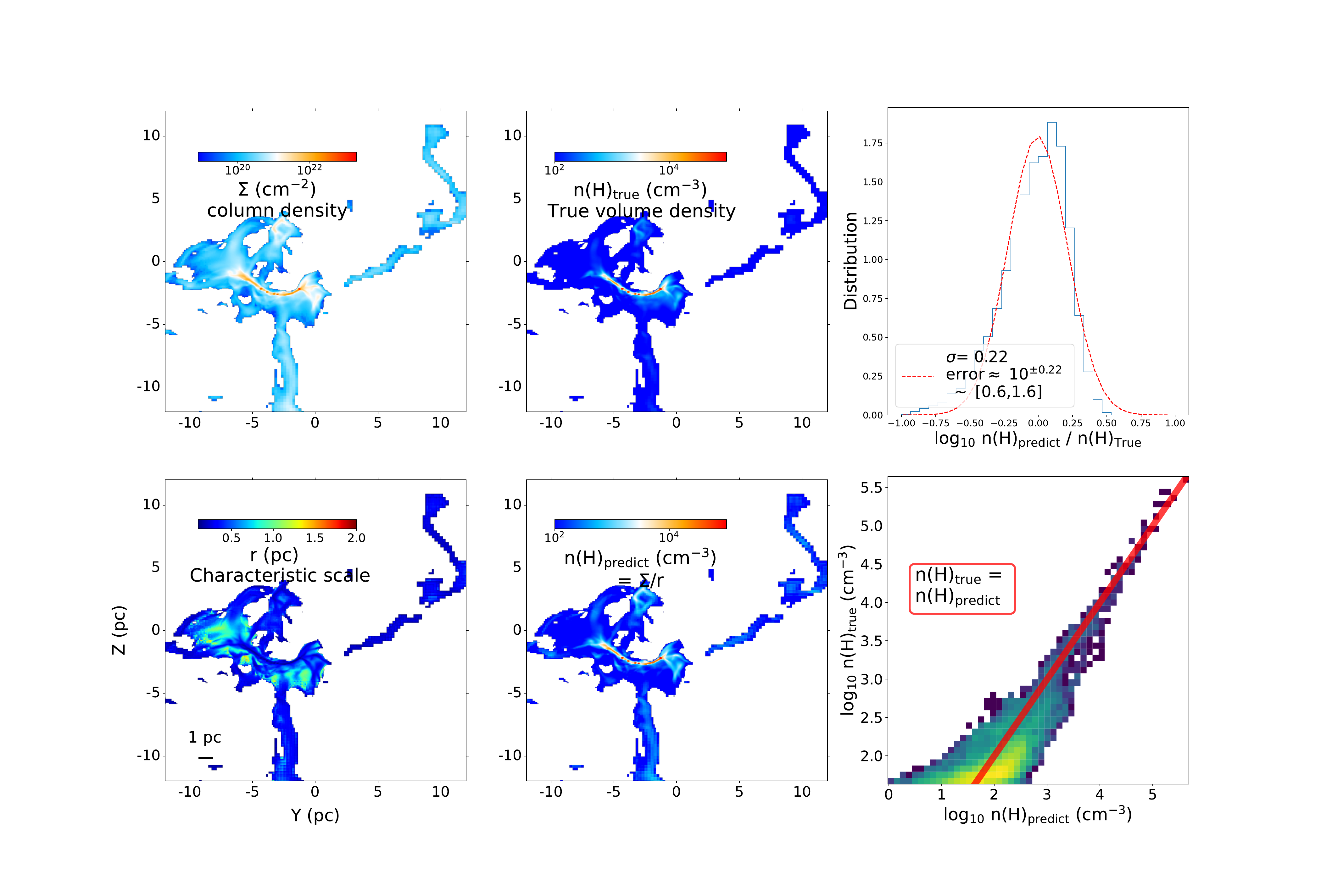}
    \includegraphics[width=0.95\linewidth]{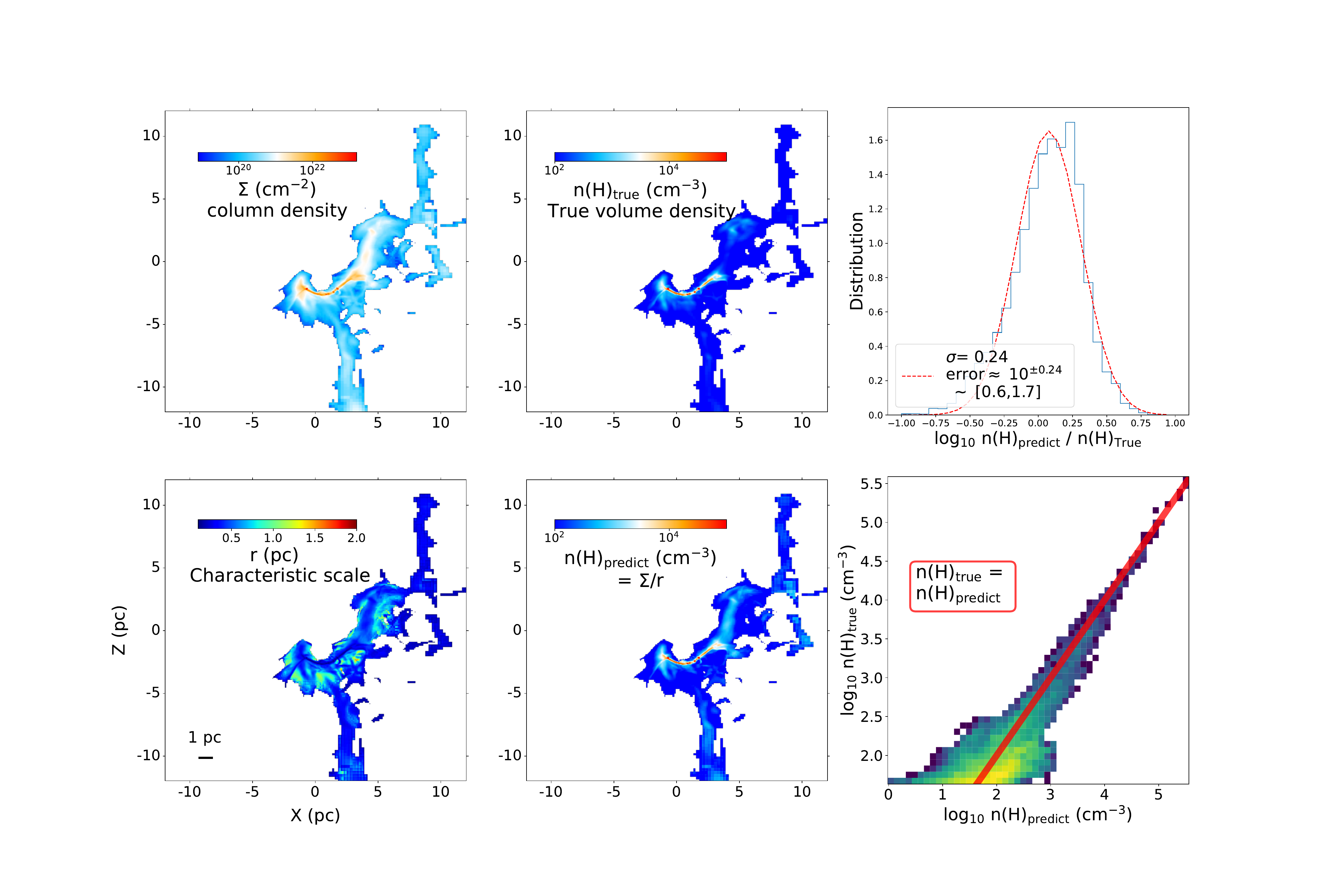}
    \caption{{\bf Validation for additional M3 projections.} 
    These panels replicate the analysis from Fig.\,\ref{figM3} for projections along the $x$- and $y$-axes.}
    \label{figM3xy}
\end{figure}

\begin{figure}
    \centering
    \includegraphics[width=0.95\linewidth]{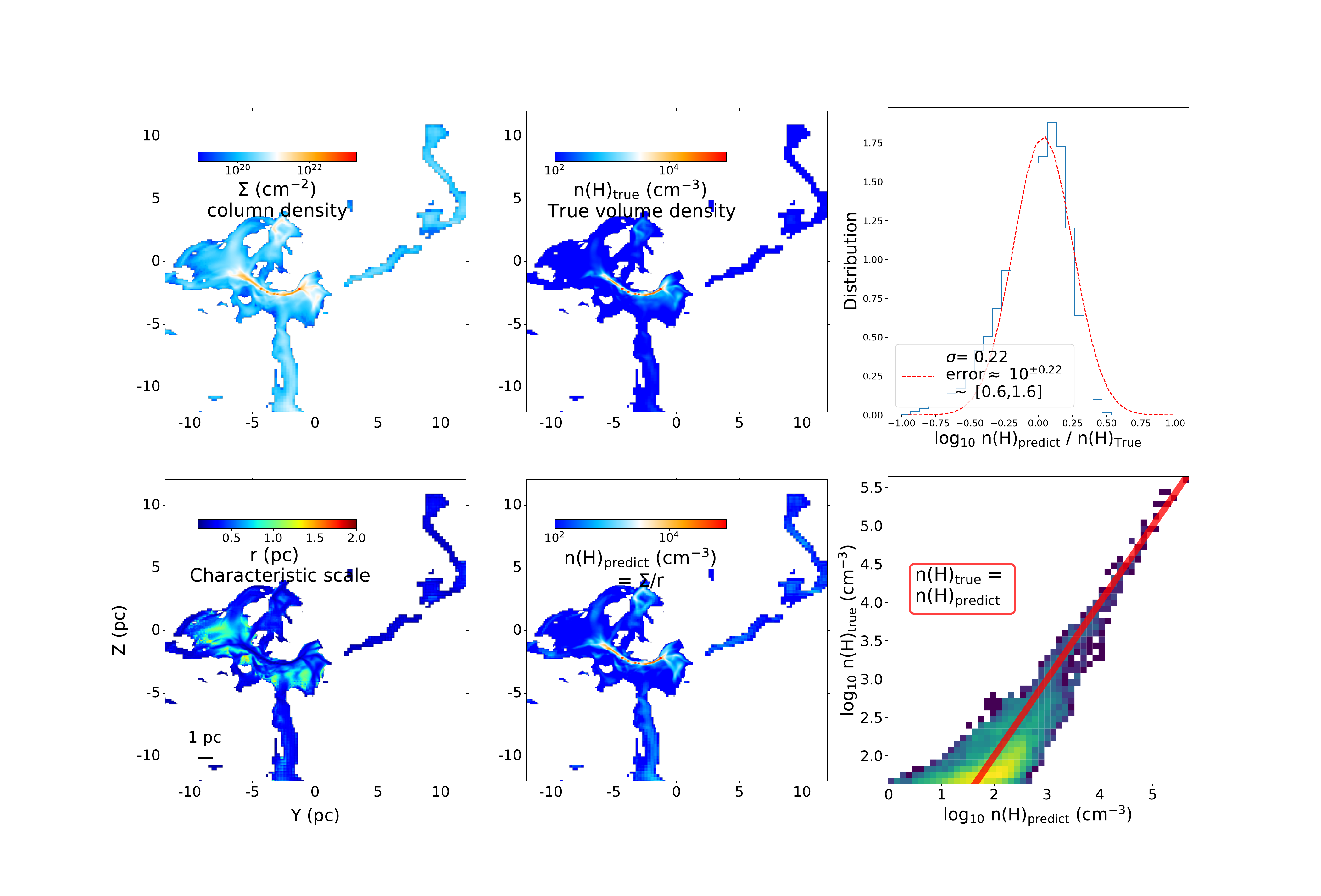}
    \includegraphics[width=0.95\linewidth]{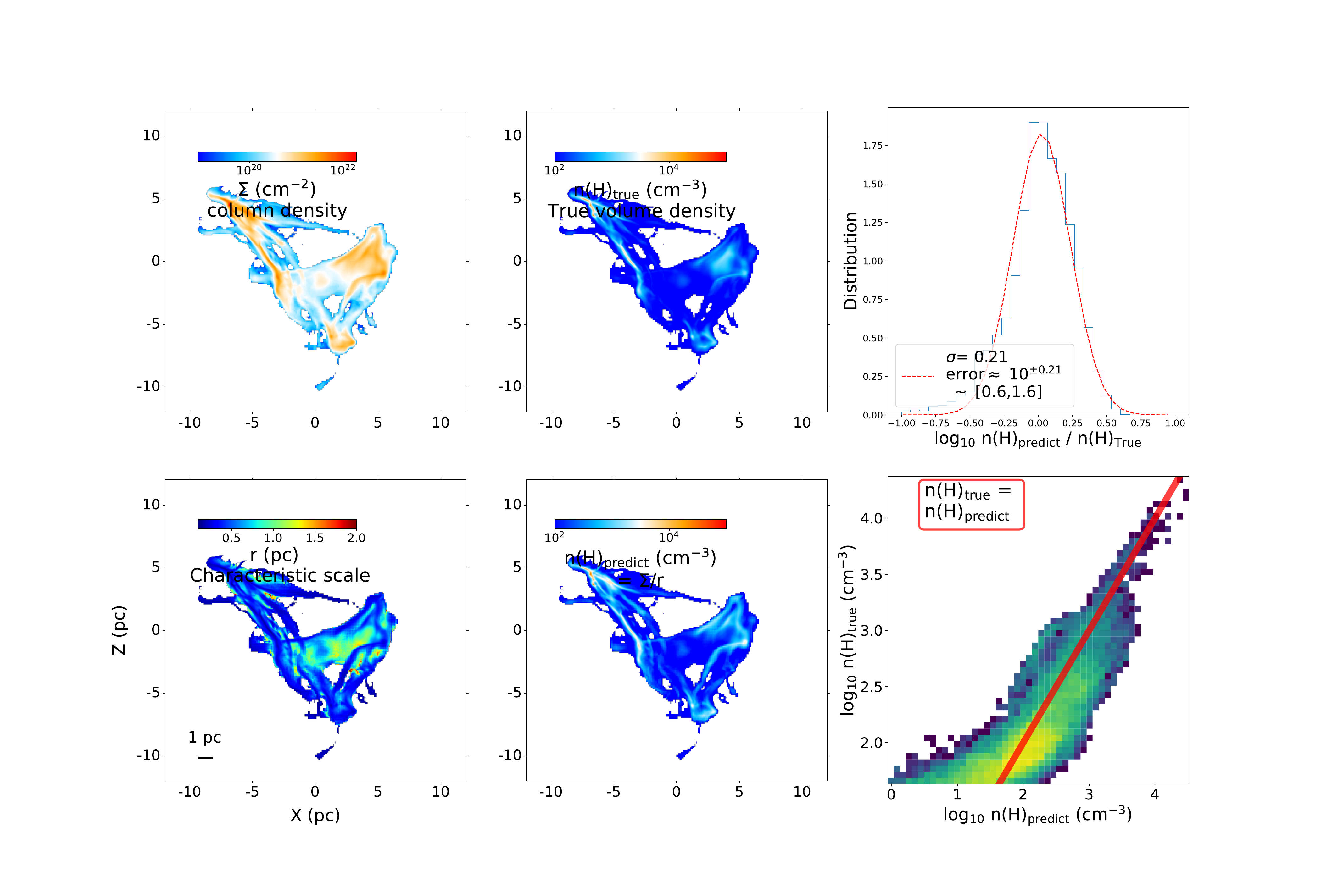}
    \caption{{\bf Validation for additional M4 projections.} 
    These panels replicate the analysis from Fig.\,\ref{figM3} for projections along the $x$- and $y$-axis.}
    \label{figM4xy}
\end{figure}

This section shows the other projected direction along the x and y-axes in cloud M3 and M4.
The predicted volume density also agree to the true volume density in 2D plane (see Fig.\,\ref{figM3xy} and \ref{figM4xy}), which does not depended on the special projected direction. 

\section{Application in Observations}\label{Ap.D}

The MDR method provides the scale and density distribution of individual molecular clouds. 
This capability supports two key analysis approaches: (1) statistical studies of entire cloud structures, or (2) targeted analysis of specific sub-regions in physical space. 

This section demonstrates MDR's broad applicability through observational examples. 
The MDR method enables direct plotting of physical parameters against scale, revealing fundamental structure-property connections, which enable diverse secondary applications, from star formation studies to turbulence analysis.

We apply MDR to ISF of Orion A, which has rich density and spectral data.
The raw data are column density fitted by Herschel continuous and the NH$_3$ spectral data observed by GBT (detailed shown in Ap.\,\ref{data}).

This section demonstrates the application of the Multiscale Density Reconstruction (MDR) method using observational data to analyze the structure and dynamics of the Orion Integral shape Filament (ISF). 
The analysis is based on column density maps derived from Herschel continuum observations \citep{2010A&A...518L...2P,2010A&A...518L...3G,2013ApJ...763...55R,2013ApJ...777L..33P} and NH$_3$ spectral line data obtained with the GBT\citep{2017ApJ...843...63F}.

\begin{figure}
    \centering
    \includegraphics[width=0.95\linewidth]{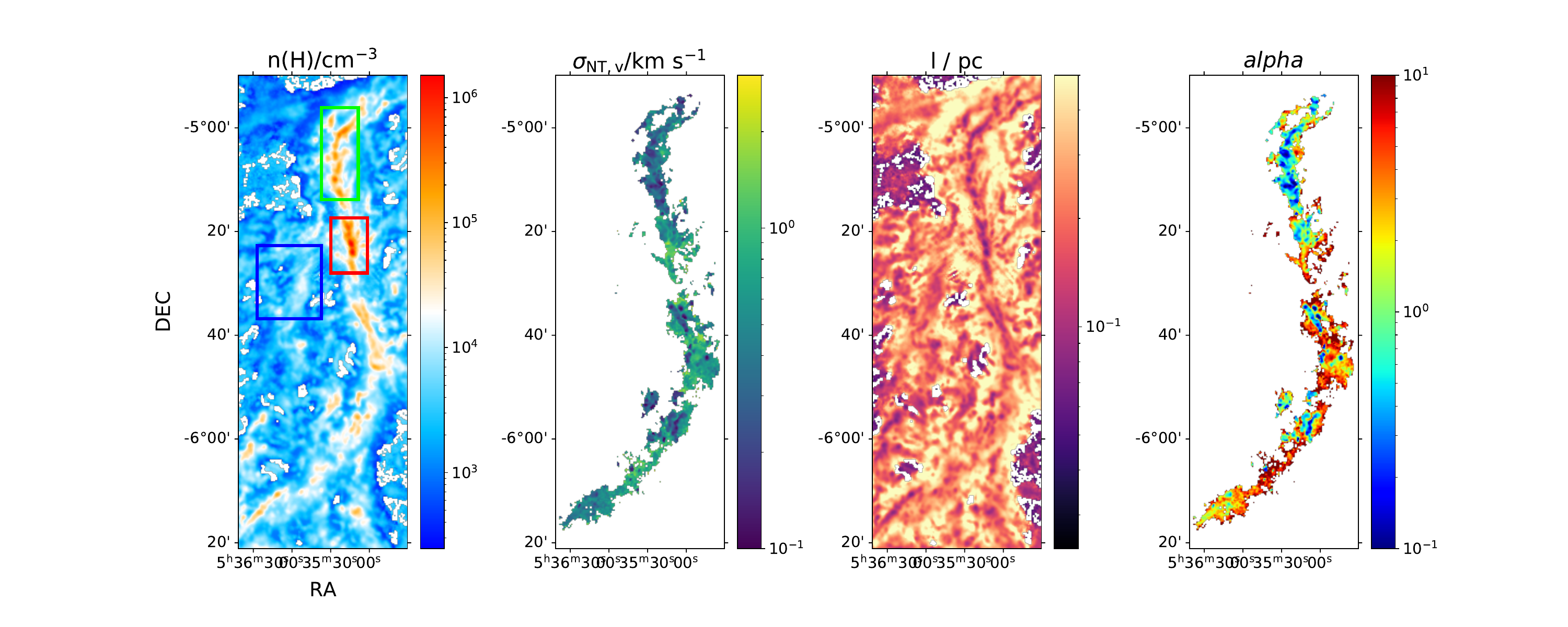}
    \caption{{\bf Mapping volume density, non-thermal velocity dispersion, characteristic scale, virial ratio in ISF.}
    The red, green, blue box in density map (left panel) show the position of OMC-1, OMC-2/3, and a part of Orion extension bubble.}
    \label{figISFs}
\end{figure}

\begin{figure}
    \centering
    \includegraphics[width=0.5\linewidth]{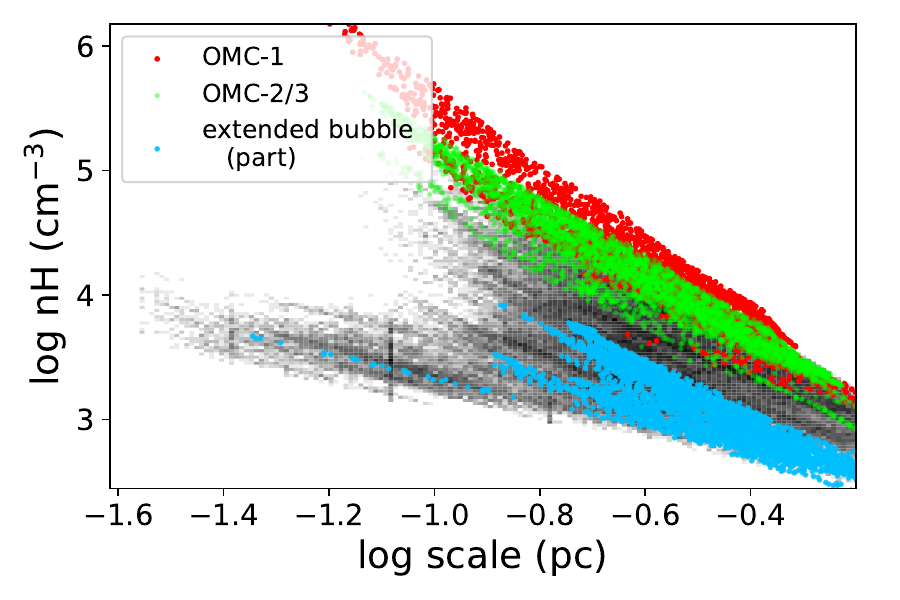}
    \includegraphics[width=0.5\linewidth]{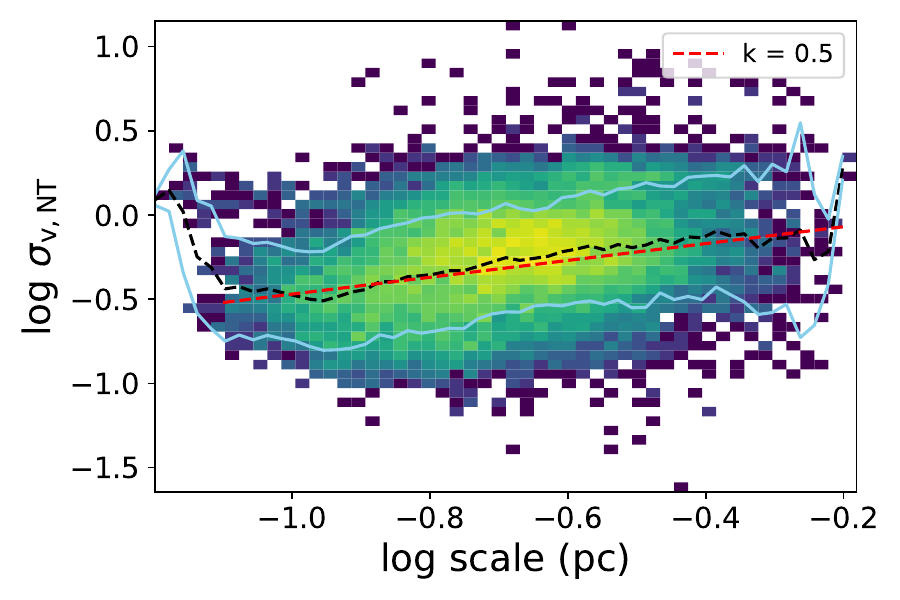}
    \includegraphics[width=0.5\linewidth]{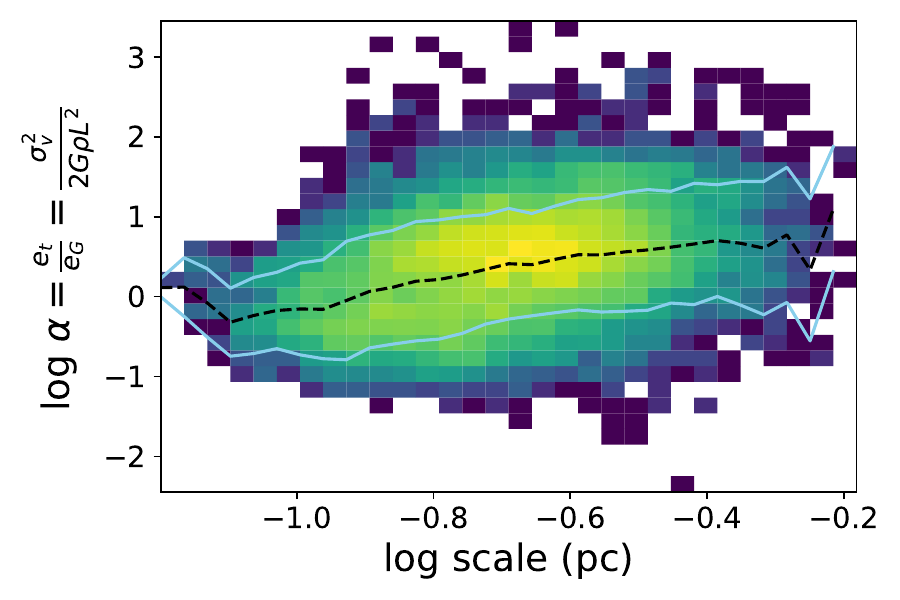}
    \caption{{\bf Distribution between scale and volume density, velocity dispersion, and virial ratio.}
    The red, green, and blue scatter in top panel, density-scale distribution show the distribution of OMC-1, OMC-2/3, and a part of Orion extension bubble, whose locations are shown in Fig.\,\ref{figISFs}.
    The black line shows the main structure, and blue lines display the dispersion region, which means the mean and dispersion values are fitting a Gaussian distribution at each given scale range.}
    \label{figISFr}
\end{figure}

The MDR method was applied to the column density data to extract the volume density n(H) and characteristic scale l (see Fig. \ref{figISFs}). 
The resulting density-scale distribution presents two distinct branches in the n(H)-l plane (Fig. \ref{figISFr}). 
Different regions within the ISF, specifically the collapsing OMC-1 region \citep{2017A&A...602L...2H}, the OMC-2/3 region, and part of an extended bubble, exhibit markedly different distributions within this plane. 
This suggests that these regions possess different density profiles and likely represent different star formation states\citep{2022MNRAS.514L..16L}.

To further study the gas dynamics, we analyze turbulent parameters using the derived scale maps and spectral line data. 
The non-thermal velocity dispersion $\sigma_{\rm v, NT}$, which approximates turbulent motions, was calculated as $\sigma_{\rm v, NT} = \sqrt{\sigma_v^2 - \frac{K_B T_{gas}}{\mu m_H}}$. 
Plotting $\sigma_{\rm v, NT}$ against scale $l$ under the assumption of isotropy (Fig. \ref{figISFr}) reveals a scaling relation within the ISF of $\sigma_v \propto r^{0.5}$. 
This trend, similar to Larson's velocity-size relation observed in individual clouds \citep{1981MNRAS.194..809L} by isotropy assumption, is consistent with previous studies \citep{2004ApJ...615L..45H} by principal component analysis.

Building on the turbulent velocity measurement, we calculated the virial ratio ($\alpha$), defined as the ratio of turbulent kinetic energy to gravitational potential energy: $\alpha = e_t/e_g = 0.5 \rho \sigma_v^2 / (G\rho^2 l^2) = \sigma_v^2 / (2 G \rho l^2)$, where the three-dimensional velocity dispersion $\sigma_{v, 3D} = \sqrt{3} \sigma_{\rm v, NT}$. 
Examining $\alpha$ as a function of scale (Fig. \ref{figISFr}) shows that within the ISF, the virial ratio decreases with decreasing scale. 
This trend describes enhanced star formation activity as the scale decreases, where gravitational binding energy dominates over turbulent energy.

\section{Consistency Check with HC$_3$N Tracers in B213} \label{Ap.E}

\begin{figure}
    \centering
    \includegraphics[width=0.5\linewidth]{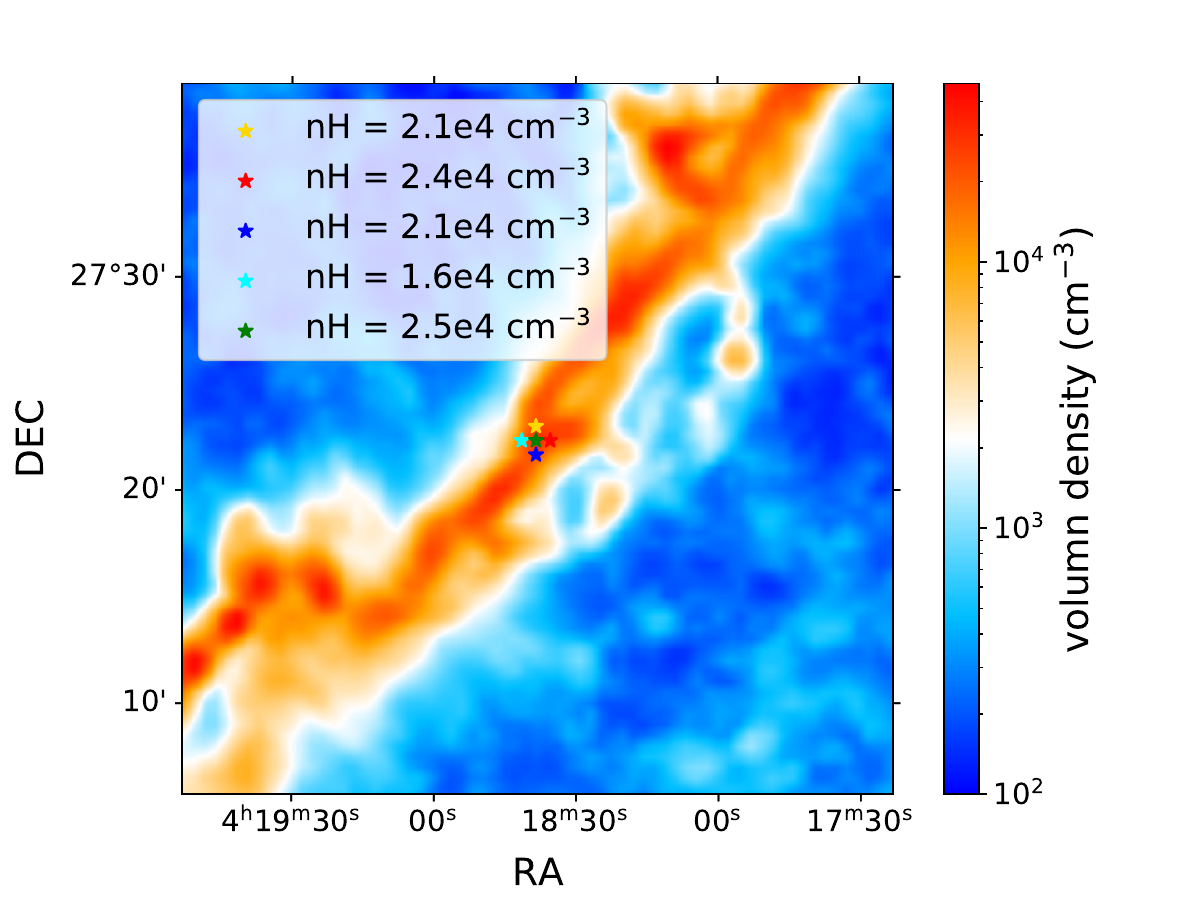}
    \caption{{\bf Predicted volume density by MDR in B213.}
    The background shows the volume density predicted MDR from column density.
    The stars are the position probed by HC$_3$N line ratios\citep{2012ApJ...756...12L}.
    The values in the label show the volume density measured by MDR.}
    \label{figB213}
\end{figure}

To provide another check under actual observations, we applied MDR to the B213 filament, where volume density estimates have been obtained using a different technique: the line ratios of the HC$_3$N molecule \citep{2012ApJ...756...12L}. 
This chemical tracer method relies on complex molecular excitation physics, different to the mathematical model, MDR.

As Fig.\,\ref{figB213} shows, the MDR-predicted volume densities at the five positions probed by HC$_3$N are in excellent agreement with the values from the chemical method. The MDR-derived value of $(2.1 \pm 0.3)\times 10^4$ cm$^{-3}$ aligns with the HC$_3$N measurement of  $(1.8 \pm 0.7)\times 10^4$ cm$^{-3}$.
This provides independent verification of MDR's applicability to actual observations.


\end{document}